\documentclass[usegraphicx,a4]{pasa}
\usepackage{color,times,hyperref}
\usepackage{verbatim} 
\newcommand{\nb}[1]{\ensuremath{^{#1}}}
\newcommand{\code}[1]{{\sc #1}}

\newcommand{\sqdeg}{\ensuremath{\mathrm{deg}^2}}

\title[The SkyMapper Transient Survey]
   {The SkyMapper Transient Survey}
\author[Scalzo et al.]
{
    R.~A.~Scalzo\nb{1,2,3},\thanks{\texttt{rscalzo@anu.edu}}
    F.~Yuan\nb{1,2},
    M.~J.~Childress\nb{1,2,4},
    A.~M{\"o}ller\nb{1,2},
    B.~P.~Schmidt\nb{1,2},
    B.~E.~Tucker\nb{1,2},
    B.~R.~Zhang\nb{1,2},
    P.~Astier\nb{2,3},
    M.~Betoule\nb{2,3},
    and N.~Regnault\nb{2,3} \\
    \affil{\nb{1} Research School of Astronomy and Astrophysics,
           Australian National University, Canberra, ACT 2611, Australia}
    \affil{\nb{2} ARC Centre of Excellence for All-sky Astrophysics
           (CAASTRO)}
    \affil{\nb{3} Centre for Translational Data Science, University of Sydney, NSW 2006, Australia}
    \affil{\nb{4} School of Physics and Astronomy, University of Southampton, Southampton, SO17 1BJ, UK}
    \affil{\nb{5} Laboratoire de Physique Nucl\'eaire et des Hautes \'Energies,
            Universit\'e Pierre et Marie Curie Paris 6,
            Universit\'e Paris Diderot \\ Paris 7, CNRS-IN2P3,
            4 place Jussieu, 75252 Paris Cedex 05, France}}

\jid{PASA}
\doi{10.1017/pas.\the\year.xxx}
\jyear{\the\year}

\usepackage[authoryear]{natbib}
\bibpunct{(}{)}{;}{a}{}{,}
\begin{document}

\begin{abstract}
The SkyMapper 1.3~m telescope at Siding Spring Observatory has now begun
regular operations.  Alongside the Southern Sky Survey, a comprehensive
digital survey of the entire southern sky, SkyMapper will carry out a search
for supernovae and other transients.  The search strategy, covering a total
footprint area of $\sim2000$~\sqdeg\ with a cadence of $\leq 5$ days, is optimised
for discovery and follow-up of low-redshift type Ia supernovae to constrain
cosmic expansion and peculiar velocities.  We describe the search operations
and infrastructure, including a parallelised software pipeline to discover
variable objects in difference imaging; simulations of the performance of the
survey over its lifetime; public access to discovered transients; and some
first results from the Science Verification data.
\end{abstract}


\maketitle

\section{INTRODUCTION}
\label{sec:intro}

The advent of automated, wide-field survey telescopes has revolutionised
astronomy by dramatically increasing the sky area that can be observed to a
given depth in a short span of time.  At the same time, automation and
digitization of the end-to-end operation of these telescopes, from routine
operation to data reduction to data storage, has produced an unprecedented
wealth of data to mine for new patterns and objects.  The surveys employing
these telescopes have created digital maps of large sky areas, such as the
Sloan Digital Sky Survey \citep[SDSS;][]{sdss}.  They also enable increasingly
intensive, untargeted monitoring of large sky areas for variable and transient
objects.  Such monitoring reduces the selection bias associated with targeting
particular sky areas or host galaxies, and results in large, homogeneous
samples of all transient phenomena in the sky in the targeted wavelength
range to a certain magnitude limit.  Completed
time-domain surveys with a footprint larger than 1000~\sqdeg\ include
Palomar-QUEST \citep{pq} and the Palomar Transient Factory
\citep[PTF;][]{ptf,ptf2}.  Ongoing wide-area time-domain surveys include
Pan-STARRS \citep{ps1}, LaSilla-QUEST \citep{lsq}, the iPTF extension to PTF,
and the Catalina Real-Time Transient Survey \citep[CRTS;][]{crts}.

The SkyMapper 1.3~m robotic telescope \citep{keller07} at Siding Spring
Observatory has commenced the Southern Sky Survey, an automated,
digital survey of the southern sky.  Alongside this survey (``Main Survey''),
the SkyMapper Transient Survey (SMT) 
is a search for supernovae and transients in the local Universe optimised to discover and follow up SNe~Ia for cosmology.

A major science driver for time-domain surveys 
is the study of the
Universe's accelerating expansion \citep{riess98,schmidt98,scp99} and
the parameters of the ``dark energy'' which drives it, through
the discovery and follow-up of type Ia supernovae (SNe~Ia). Contemporary cosmological analyses such as the Joint Lightcurve Analysis \citep[JLA;][]{betoule14} require both high-redshift and low-redshift SNe~Ia to make inferences about the dark energy equation of state; the low-redshift SNe~Ia mainly constrain the mean absolute magnitude of SNe~Ia, while the high-redshift SNe~Ia use luminosity distances to map the Universe's scale factor over cosmic time.  Presently the high-redshift SNe~Ia sample is composed of magnitude-limited surveys such as the Supernova Legacy Survey \citep[SNLS;][]{sullivan11, conley11} and the ongoing Dark Energy Survey \citep[DES;][]{des}. Meanwhile the nearby supernova sample comes from a myriad of surveys through the 2000's such as the Harvard-Smithsonian Center for Astrophysics (CfA) surveys (CfA1-4; \citep{cfa,cfa2,cfa3,cfa4} and the Carnegie Supernova Project \citep[CSP;][]{csp}, which follow up SNe discovered in the automated Lick Observatory Supernova Search
\citep[LOSS;][]{loss,loss2} or by amateur astronomers. These supernovae commonly were found by targeting large nearby galaxies. The accuracy and precision with
which cosmological parameters can currently be measured from SNe~Ia are limited by
systematic errors, particularly photometric calibration
\citep{betoule13,betoule14}, but also including uncertainties in dust
extinction \citep{phillips13,scolnic14,burns14}, potential population
diversity \citep{qhw07,wang09,kelly10,sullivan10,childress13,kelly15}, evolution \citep{kim04,howell07,sullivan09,milne15}
over a range of redshifts, or the influence of peculiar velocities \citep{davis11} including coherent bulk flows \citep{hg06}.

Of the aforementioned systematics, the contributions to the uncertainty budget from photometric calibration and peculiar velocities are magnified by the heterogeneity of the low-redshift SN~Ia sample, in particular by the numerous telescopes used to observe them, and by their non-uniform spatial distribution. 
SkyMapper aims to address these limitations by searching a wide sky area uniformly with a short ($\leq 5$ days) cadence, in multiple well-determined bandpasses~\citep{bessell11}. Thus the resultant low-redshift ($z < 0.1$) SN~Ia sample will be well-calibrated and magnitude-limited, with a more similar selection function to the high-z sample. SMT will therefore be very
useful for measurements of cosmic expansion and peculiar
velocities associated with bulk flows and cosmic structure, and for studies of
type Ia supernova physics aimed at improving SNe~Ia as distance indicators. SkyMapper is also unique in its spatial overlap with the DES footprint, positioning SMT as an optimal low-redshift anchor.

Here we introduce the infrastructure and operations of the SkyMapper Transient Survey,
and present some first results including performance during an early Science
Verification period. The SN Survey has begun operating at scale from
April 2015, and has released candidates and classifications to the public. This paper will be followed shortly by an early data release of $\sim 30$ SNe~Ia to date, and by individual papers on peculiar transients such as superluminous supernovae (\S\ref{sec:weirdo}). The structure of the paper is as follows: \S 2 gives an overview of SkyMapper telescope and the Main Survey. \S 3 describes the search pipeline and follow-up procedure, while \S 4 focuses on survey strategy. We discuss and evaluate the performance thus far in \S 5, and present some early results in \S 6.

\section{THE SKYMAPPER TELESCOPE AND MAIN SURVEY}
\label{sec:ops}

The SkyMapper telescope, and the Main Survey infrastructure and science goals,
are described in detail in \citet{keller07}; we briefly summarise the most
relevant details below. 

\newcommand{\sex}{\code{SExtractor}}
\newcommand{\swarp}{\code{SWarp}}
\newcommand{\hotpants}{\code{hotpants}}




SkyMapper is a 1.3~m, $f/4.8$ telescope at Siding Spring Observatory, operated
by the Australian National University on behalf of the Australian astronomical
community.  The telescope has a 5.7~\sqdeg\ field of view, covering a square
$2.4~\deg~\times~2.4~\deg$ area with a fill factor of 91\%.  The 268-Mpix
imager has a pixel scale of 0.5~arcsec/pix.  Available filters include
SDSS-like \emph{griz}, Stromgren \emph{u}, and a custom-made, intermediate-band
\emph{v} filter specific to SkyMapper \citep{bessell11}.  The \emph{v} filter covers the
range 3670--3980~\AA, to allow simultaneous measurements of surface gravity
and metallicity from broad-band photometry; it is optimised to enable the Main
Survey's key science goals in galactic archaeology, particularly the
identification of extremely metal-poor stars \citep[e.g.][]{keller14}.

The calibration of the Southern Hemisphere standard stars in the SkyMapper
photometric system is accomplished through a ``Short Survey'' of images
with short (5--10~sec) exposure times, targeting secondary standard stars
with magnitudes between 8.5 and 15.5 \citep{keller07}.  This part of the
survey requires at least three images of each part of the sky in all six
SkyMapper filters, under photometric conditions.  Photometric superflats
are also formed by taking a series of dithered exposures, moving a standard
star across the surface of the mosaic, to characterise and remove patterns
of scattered light.  The absolute zeropoint of SkyMapper will be determined
from stars in the Walraven photometric system \citep{walraven} with Hubble
Space Telescope spectrophotometry from the Next Generation Spectral
Library.\footnote{
{http://lifshitz.ucdavis.edu/~mgregg/gregg/ngsl/ngsl.html}}
Changes in the SkyMapper optical
throughput will be monitored using a photodiode system \citep{stubbs06,sndice}.

SkyMapper's routine observing is fully automated.  A schedule is input for each
night as a list of observation requests, each comprising a desired bandpass,
exposure time, and optimal window during which the observation can take place.
Automated safety measures will close the dome in case of cloud or inclement
weather.  The Main Survey nightly schedules are determined each night
according to a survey strategy which balances its various science goals,
but also monitors the observing conditions
(for example, seeing or photometricity) to trigger execution of third-party
programs designed to take advantage of conditions unsuitable for Main Survey
operations.  The SN Survey is an example of such a third-party program.
Observations can thus be scheduled in ``classical'' mode, occupying specific
blocks of time, or in ``queue'' mode to respond to changing observing
conditions.


\section{SKYMAPPER TRANSIENT SURVEY PIPELINE}

Although the SN Survey and the Main Survey both use the SkyMapper telescope,
the SN Survey has its own data reduction pipeline infrastructure, maintained
and run separately from the Main Survey data pipeline.  This pipeline enables
additional processing (e.g. image subtraction) beyond Main Survey requirements, ensures rapid ($< 12$~h) turnaround for discovery of new
transient candidates, and provides additional data to support situational
awareness of active candidates, such as historical light curves, a web
service enabling follow-up of transients, and annotations by users about the transient type and characteristics.

\subsection{Image subtraction workflow}

\begin{figure*}
\begin{center}
\resizebox{0.9\textwidth}{!}{\includegraphics{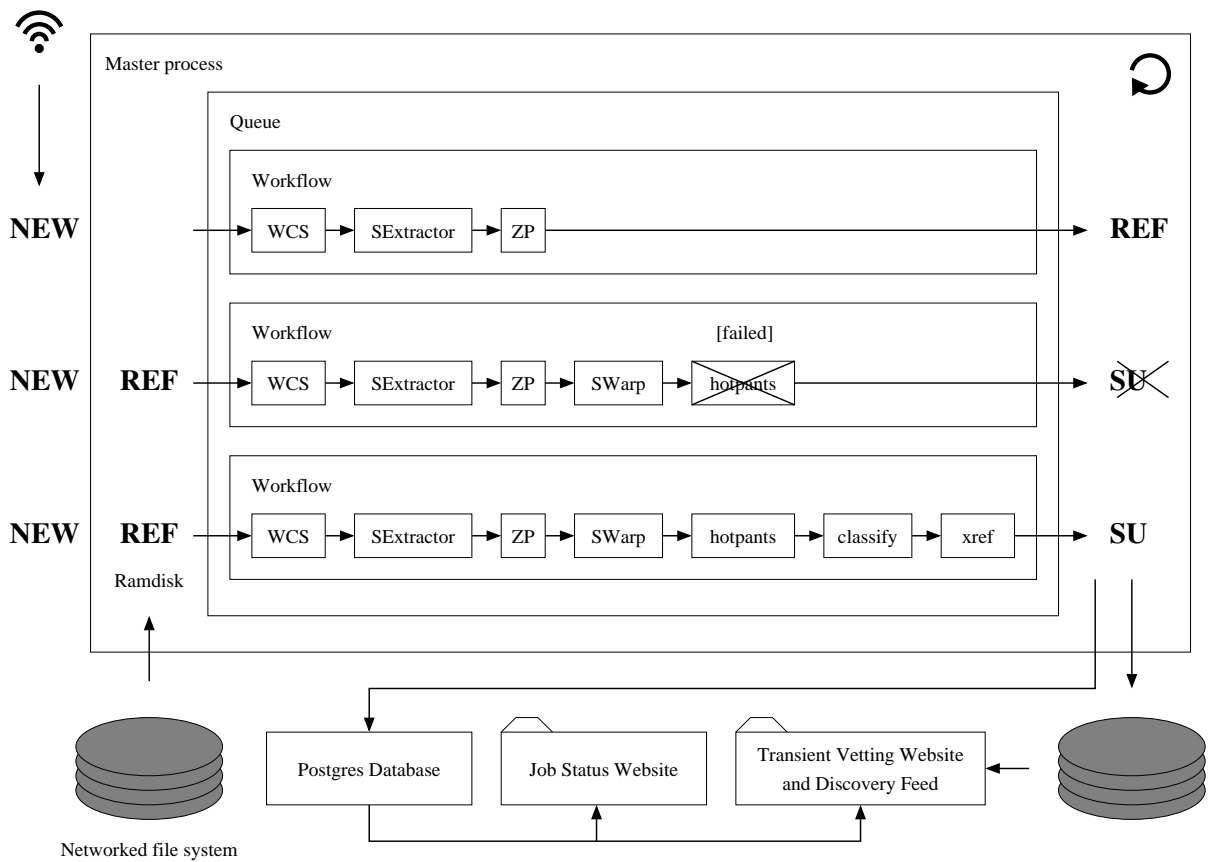}}
\end{center}
\caption{Architecture of the SkyMapper SN Search pipeline.}
\label{fig:subpipe-flow}
\end{figure*}

The image subtraction pipeline is written almost entirely in Python,
with some C++ extensions for pixel-level image processing (e.g. flat fielding)
and incorporating commonly used open-source modules  \citep[e.g.,][]{sex}
where available.  The stages of a particular workflow are shown in
Figure~\ref{fig:subpipe-flow}.
Image subtraction for the SN Survey requires a pre-existing template image of the sky (``REF'') to remove host galaxy light and non-variable sources from each new exposure (``NEW'').  Image subtraction involves astrometric
resampling and rescaling of the REF to match the NEW, and convolution of the
REF by a suitably chosen kernel so that its point-spread function (PSF)
matches that of the NEW.  During Science Verification, a cache of REF images
was built up automatically; now, if no REF image is available in a given part
of the sky to subtract from a NEW image, the NEW image is simply added to
the cache as a REF.  We require that each REF image have a narrower PSF than the NEW image, so that the REF is always convolved to the NEW; this minimises
correlations between pixels associated with newly discovered SNe.  We also
require that the REF image be taken at least two weeks prior to the NEW image,
since the rise time of a typical SN~Ia to maximum light is about 17~days.
Following the start of Main Survey operations in
April 2015, the SN Survey began coordinating with the Main Survey
to use completed Main Survey exposures as REF images for template subtraction.
This ensures that suitably deep REF images exist in all four SN Survey
filter bands (\emph{vgri}) before beginning to search a given field, so that
follow-up light curves can be generated immediately in any filter.

\begin{figure}
\begin{center}
\resizebox{0.5\textwidth}{!}{\includegraphics{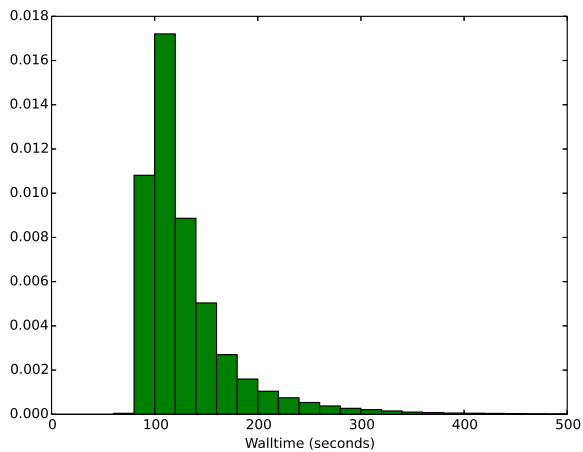}}
\end{center}
\caption{Distribution of processing times for successful subtraction jobs.}
\label{fig:walltime}
\end{figure}

Reduction of a SkyMapper SN Survey exposure begins with a simplified version
of the Main Survey pipeline workflow described in \citet{keller07}.
The mosaic images are split into individual $2048 \times 4096$ CCD images.
An overscan region is subtracted from each half of the CCD (corresponding to
different amplifiers).  Flat fields are constructed nightly from dome flat
images and applied to science exposures after overscan subtraction.
A bad pixel map is created to flag pixels based on consistent deviation from
a reasonable gain range and/or erratic behavior.  A quick large-scale
astrometric solution for the mosaic is produced using \code{astrometry.net}.
If the astrometric registration fails (e.g., because of heavy cloud), or if
the image quality is poor (FWHM $> 4$~arcsec, or elongation $> 1.2$),
the image is discarded at this stage and undergoes no further analysis.
For each individual CCD of a NEW exposure, the workflow then proceeds
through the following stages:
\begin{enumerate}
\item {\bfseries WCS:} The world coordinate system for the NEW is refined,
      with higher-order distortions described in the zenithal polynomial
      (ZPN) representation.  This produces astrometry accurate to about
      0.1~arcsec.
\item {\bfseries SExtractor:}  Sources are detected in the NEW image with \sex\
      \citep{sex} and aperture photometry is extracted over a series of
      apertures.
\item {\bfseries ZP:}  A preliminary photometric zeropoint is produced by
      comparison to APASS \citep{apass}.  Analysis of data from June 2014
      suggests that the mean color terms for transforming between the APASS
      and SkyMapper \emph{gri} filters are small ($<0.05$).  The SN Survey
      will eventually be tied to the same photometric system as the
      Main Survey, pending completion of Main Survey fields in the search
      area.  If an image has no corresponding REF, it is added to the REF
      cache at this stage and no further processing takes place.
\item {\bfseries SWarp:}  The REF image is resampled to the coordinate
      system of the NEW image using \swarp\ \citep{swarp}.
\item {\bfseries hotpants:}  The REF is scaled to the NEW flux level,
      convolved with a spatially-varying kernel to match the NEW PSF as
      accurately as possible, and subtracted from the NEW using
      \hotpants\footnote{{http://www.astro.washington.edu/users/becker/v2.0/hotpants.html}}
      to produce a subtracted image (``SUB'').  Sources are detected in the
      SUB image with \sex.  Each SUB image inherits the world coordinate
      system and photometric zeropoint of the corresponding NEW image.
\item {\bfseries classify:}  All detections on the SUB are run through an
      automated classification routine (see \S\ref{subsec:classify})
      to determine the likelihood that they are real astrophysical sources
      rather than artifacts from an imperfect subtraction process.
\item {\bfseries xref:}  All high-quality detections in the SUB image are
      astrometrically matched to previous detections.  For sources
      passing a threshold number of high-quality detections in one or more
      subtractions, a historical light curve is compiled using all
      detections of the transient at that position.
\end{enumerate}
Figure \ref{fig:walltime} shows the distribution of ``wall times''
for the processing of a single SkyMapper CCD (running on a single core).
The median processing time is 115 seconds from initial reduction to
automatic flagging of candidates.  While this is slower than real-time, it is
fast enough to process an entire night's worth of exposures in less than
24~hours, allowing the pipeline to keep up with the flow of data.
During Science Verification, we found that the end-to-end success rate for
subtraction jobs was close to 99\%, with a small number of failures easily
traced to low-quality input data
(due, for example, to poor weather conditions).
\subsection{Image subtraction pipeline job control}
The pipeline runs on a custom-built cluster named  Maipenrai\footnote{{http://www.maipenrai.com.au}} hosted at the
Australian National University's Research School for Astronomy and
Astrophysics (RSAA) in Canberra.  The cluster has 48 cores with 192~GB of
random-access memory and 44~TB of network-mounted disk space.  A small part
of available memory (32~GB) is set aside as a fast virtual file system
(``ramdisk'').  This ensures that I/O-intensive processing by third-party
image processing programs can be performed directly in memory without
modifying the code, dramatically improving performance.
Relational information about images, pipeline jobs, and transient objects
discovered is hosted in a Postgres database, accessed through the Django
web framework\footnote{{https://djangoproject.com}}.  Commonly used catalogs, such as UCAC2
\citep{ucac2}, 2MASS \citep{2mass}, and APASS \citep{apass}, are accessed
via a separate Postgres database hosted locally at RSAA.
Figure~\ref{fig:subpipe-flow} also shows a schematic representation of the
flow control for the pipeline.  To produce an architecture that is efficient,
fault-tolerant, and transparent, we adhere to the following
design principles:
\begin{enumerate}
\item A master process coordinates assignment of jobs to up to 32 worker
      processes at a time, monitoring their state through a polling loop.
      Each SkyMapper exposure is reduced using one worker process per
      $2048 \times 4096$ CCD, so that individual images fit easily in memory
      and no special software for reducing large mosaic images is needed.
\item Only the master process is allowed to transfer inputs and outputs
      of worker processes between shared disk and the ramdisk, which it does
      synchronously at the beginning and end of a polling cycle.
      This prevents many worker processes from accessing shared disk at
      once, placing minimal strain on networked file systems.
\item Although worker processes can query databases, the master process is
      responsible for updates to database tables, aggregating results from
      various processes to minimise I/O overhead.
\item The SN Survey layout is organised around a set of fixed fields on
      the sky, corresponding to a subset of the Main Survey fields.
      All shared disk storage is organised into subpaths corresponding to
      unique field/filter/CCD combinations.
\item Each worker process runs through a modular workflow, logging both the
      system calls needed to execute particular steps and the output of
      those steps to a log file.  Any step which fails can be rerun easily
      based on the logged system calls, speeding up debugging.  Process
      status and log files can be accessed quickly through a web interface.
\end{enumerate}


\subsection{Selecting candidates for photometric follow-up}
\label{subsec:classify}
The astrometric, photometric, and PSF matching of the REF to the NEW will
in general not be perfect.  Image subtraction artifacts not corresponding
to astrophysical variable objects are easily recognizable to the human eye
as anything in the SUB image not resembling a point source.  However,
these artifacts are much more numerous, outnumbering true astrophysical
variable objects by more than an order of magnitude even in relatively
clean subtractions.  Initial triage of detected objects on subtractions must
therefore be automated.

To address this challenge, we have implemented a series of machine learning
classifiers to distinguish ``Real'' astrophysical objects from ``Bogus''
artifacts or cosmetic features appearing in the search images.
Our first version of this classifier used the Python-based machine learning
package \code{milk}\footnote{{https://github.com/luispedro/milk}} to implement
a random forest classifier modeled after \citet{bloom12}.  For more recent
versions, we have switched to the random forest implementation in \code{sklearn}
\citep{sklearn}, which trains more quickly and makes cross-validation easier.

At all stages of its development, the performance of the classifier has been
limited mainly by the availability of training examples of Real SNe.
The first version of the classifier used training data from early SkyMapper
commissioning (August 2011), based on a sample of detections
visually scanned and tagged as visually similar to Real or Bogus detections
by human scanners.  When evaluated against Real SN detections from contemporary
data, this version of the classifier performs no better than random chance ---
possibly due to the lack of confirmed Real supernovae in the training set,
and to the dramatic changes in the SkyMapper PSF from commissioning through
to current operations.  After the Zooniverse campaign in March 2015,
we retrained the classifier on a larger sample of Real detections
of supernovae discovered by the pipeline (see Table~\ref{tbl:sne}),
supplemented by a random selection of asteroids of varying magnitude
as examples of Real objects visually resembling (hostless)
supernovae in single exposures.
A third version was trained in October 2016 using Real discoveries from
the first year of full-time operations.  As the number of Real supernovae
increased, successive retrainings have reduced our dependence on non-supernova
detections tagged as Real, producing progressively more accurate results.

We evaluate the performance of all classifier models using
$k$-fold cross-validation, in which the data are divided into $k$ disjoint
subsets, with $k-1$ subsets reserved for training and the final subset used
for validation.  This technique enables most of the data to be used for
training while determining the impact of certain subsets of data on
classifier robustness.  We chose $k=5$ for our training.  To make a fair
estimate of the generalization error from our small sample of Real supernovae,
we placed multiple detections of the same supernova in the same fold.
This ensures that the training accounts for variations in observing
conditions and supernova magnitude, while the uncertainty in SN performance
fairly reflects variations in host galaxy background and contrast, to which
our classifier will be vulnerable when evaluating new detections.
The total dataset for cross-validation includes 688 Real detections of
57 supernovae of all spectral types, 1351 Real detections of asteroids,
and 4479 randomly selected Bogus detections; the 

\begin{figure}
\begin{center}
\resizebox{0.5\textwidth}{!}{\includegraphics{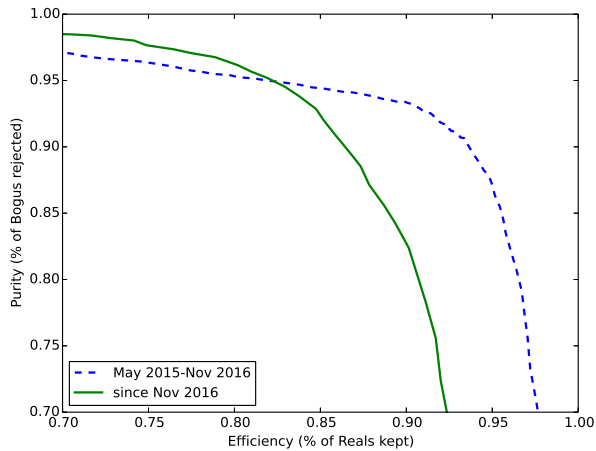}}
\end{center}
\caption{ROC curve (averaged among folds) for Real/Bogus classifier results.}
\label{fig:rb-roc}
\end{figure}

\begin{figure}
\begin{center}
\resizebox{0.5\textwidth}{!}{\includegraphics{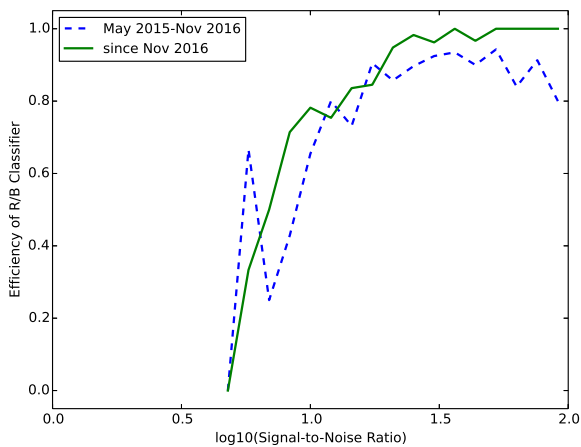}}
\end{center}
\caption{Classifier efficiency as a function of detection signal-to-noise.}
\label{fig:rb-eff}
\end{figure}

We evaluate the classifier's performance according to the efficiency
(1.0 minus the missed detection rate) and purity
(1.0 minus the false positive rate) of the classified candidates.
Figure~\ref{fig:rb-roc} shows the Receiver Operating Characteristic
(ROC) curve of these measures against each other, averaged over folds,
demonstrating the trade-off resulting by varying the score threshold
separating the Real and Bogus classes.  The more recent classifier
version is more efficient at high purity, with about 70\% efficiency
at 99\% purity (somewhat worse than the \citet{bloom12} classifier
on which it is based).  Figure~\ref{fig:rb-eff} shows the efficiency
of the more recent classifier version as a function of signal-to-noise
ratio of the detection, demonstrating recent improvements in the
effective signal-to-noise threshold and retention of bright detections.
All versions of the classifier are archived and labelled so that the
selection function for candidates can be reconstructed for later
studies (e.g., SN rates or SN Ia cosmology).

To further reduce the rate of false positives, we also require at least two
Real detections of an object at the same location on separate nights or filters.  Objects
passing this cut are astrometrically matched to existing catalogs, including
APASS \citep{apass}, the 2MASS Point Source Catalog \citep{2mass}, and the
SkyBot virtual observatory service for asteroids \citep{skybot}.
Any match better than 1~arcsec to the position of a known
point source will cause the classification of that source to be carried over
to the new candidate.

Events with at least two detections (in any band) are passed on
with annotations to astronomers to be reviewed for potential follow-up after
every night of observing.  The historical light curve is available for review,
showing photometric detections and upper magnitude limits throughout the
recent history of observations of each field.


\subsection{Follow-up}\label{section:follow-up}

Once a candidate is selected for follow-up, it is placed in a queue for
intensive monitoring by the SkyMapper telescope on a nominal 4-day cadence for \emph{gri} (5-day for \emph{v}), 
to ensure high-quality post-detection light curves. 

Spectra for classification and scientific follow-up are taken as part of the ANU WiFeS SuperNovA Program (AWSNAP; submitted to PASA), using the WiFeS integral field
spectrograph \citep{wifes} on the RSAA 2.3-m telescope at Siding Spring Observatory, and as part of other spectroscopic surveys such as
the Public ESO Spectroscopic Survey of Transient Objects \citep{pessto} and the Las Cumbres Observatory Global Telescope Network \citep{lcogt}. Follow-up targets are shared with PESSTO via a live feed, and as of mid-2016 WiFeS observations take place in Target of Opportunity (ToO) mode. An API to report SMT transient candidates through the Transient Name Server\footnote{{https://wis-tns.weizmann.ac.il/}} is already implemented and working since December 2016. Transients reported can be also found in the SMT webpage \footnote{{http://www.mso.anu.edu.au/skymapper/smt/}}.


\section{SURVEY STRATEGY AND SIMULATIONS}
\label{sec:sims}

Here we discuss the survey strategy for the SkyMapper SN Survey, which serves to maximise
the number of well-sampled SN~Ia light curves that can be included in a cosmology
sample.


\subsection{Search and follow-up strategy}

The SkyMapper SN Survey strategy is tuned to
discover SNe~Ia in the local universe ($z < 0.1$), uniformly distributed
in solid angle at high galactic latitudes ($b > 30$), and to produce 
high-quality multi-band light curves for cosmology.  The strategy includes
two components:
\begin{enumerate}
\item \emph{Rolling search mode:}  The telescope observes on a regular
      cadence ($\leq 5$~days) in the SkyMapper \emph{gr} bandpasses.
      This mode does not explicitly target known galaxies, in order to
      produce a selection function as similar as possible to high-redshift
      SN~Ia surveys such as SNLS. 
\item \emph{Follow-up mode:}  The telescope follows up fields with active
      supernovae using a tighter cadence with \emph{griv} bandpasses (see Section \ref{section:follow-up}).
      Although follow-up does not require SkyMapper's wide field, it ensures that a
      uniform accurate calibration applies both to pre-discovery photometry
      from the rolling search and to the follow-up photometry.  This mode
      also enables the SN Survey to trigger on supernovae found in other
      public supernova searches, to boost statistics and enable
      cross-calibration of SkyMapper photometry with photometry from other
      groups.
\end{enumerate}

\begin{figure}
\begin{center}
\resizebox{0.4\textwidth}{!}{\includegraphics{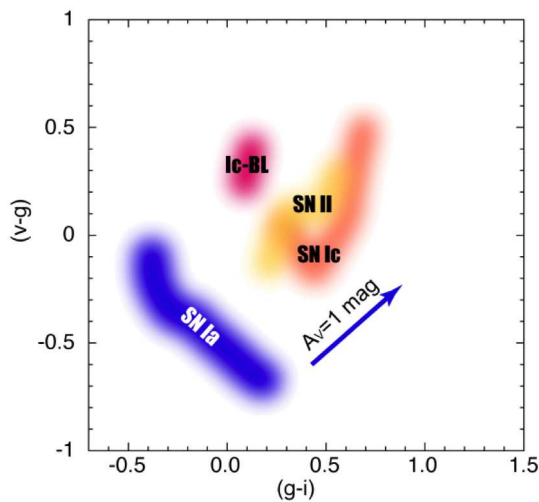}}
\caption{Subclasses of supernovae, including Type Ia and core-collapse (normal and broad-line Type Ic and Type II) in a \emph{v-g}--\emph{g-i} colour--colour plot. This figures shows that \emph{v}-band lightcurve points provide colour information for photometric selection of candidates to complement spectroscopic classification.}
\label{fig:vselect}
\end{center}
\end{figure}

The exposure times for each bandpass are constrained by the desired limiting magnitude of our survey of $\sim 20.5-21$. The chosen exposure times are 100~sec in \emph{g} and \emph{r}, 300~sec in \emph{i}, and 500~sec in \emph{v}. Early SkyMapper \emph{v}-band exposures are potentially valuable for
photometric discrimination between SNe~Ia and other types of supernovae (see Figure~\ref{fig:vselect}),
and for examining the influence
of progenitor metallicity on SN~Ia luminosities. The \emph{v}-band is read-noise
dominated for exposures less than about 500~sec long.  We therefore include
the \emph{triggered follow-up} of 500~sec \emph{v}-band observations after the
discovery of each SN~Ia in our total observing time budget.

SMT uses bad-seeing conditions which are less useful
Main Survey imaging, as it involves only detection of bright
point sources in relatively sparse fields.  Our program uses the worst
25--30\% of seeing conditions for observations in the transient search, corresponding to a threshold seeing of $> 2.3''$. This is supplemented by time in any conditions (including good-seeing) to maintain a
semi-regular cadence on active SN fields in follow-up mode. 

The number of fields covered by the footprint is constrained by the desired
cadence, depth, and wavelength coverage, as well as the total amount of
telescope time available to the survey.  To better understand these
trade-offs, we are carrying out additional simulations which will take
into account the characteristics of the instrument
(throughput, PSF, read-out noise sky background for each filter)
and historical variations in weather conditions at Siding Spring Observatory. 


\subsection{Survey geometry}

Coherent peculiar motions in the local Universe produce spatially correlated
deviations in peculiar velocities from a uniform Hubble flow; accurate
constraints on bulk flows require coverage over a large area on the celestial
sphere \citep{hg06}.  Unless the entire sky is covered uniformly,
survey geometry may affect the final performance of the survey.
\citet{haugbolle07} argued that accurate measurements of peculiar velocities
required a survey geometry that minimised the size of holes in the footprint.
The large dust extinction in the plane of the Galaxy constrains the survey
geometry, since Milky Way dust is the second-largest source of systematic
uncertainty in SN~Ia distances (after photometric calibration) in contemporary
SN~Ia Hubble diagrams \citep{conley11,betoule14}.

The influence of different survey geometries on bulk flow constraints for
the SkyMapper SN Survey has been simulated in Scrimgeour et al.
(in prep) including random sets of fields selected uniformly in area,
``glass'' geometries meant to minimise
holes in coverage, and geometries avoiding the Galactic plane
(with a maximum Milky Way extinction or minimum Galactic latitude).
They found that the total number of SNe~Ia discovered and the combination of
the SkyMapper SN~Ia sample with northern-hemisphere samples (such as PTF) were
each more influential factors than the choice of any specific survey geometry.
Therefore, it makes sense for the SkyMapper SN survey to choose fields to
minimise Galactic extinction, although new fields may be added as the survey
progresses, in order to improve constraints on the bulk flow.

The SkyMapper SN Survey will therefore concentrate on a set of low-extinction
fields ($E(B-V)_\mathrm{MW} < 0.05$). This strategy will ensure the pre-existence
of deep galaxy references, which dramatically increase the expected SN~Ia
yield in our simulations (by nearly a factor of 2) relative to the Science
Verification case where galaxy reference images are comparable in depth to
the search images.

Additionally, we are following areas of the sky currently prioritised for Main Survey coverage by other extra-galactic programs, including the Shapley Supercluster and the
footprint of the Kepler Extra-Galactic Survey (KEGS), which is using the Kepler K2 mission to monitor supernova fields at a very high-cadence of 30 minutes.  K2 fields 1, 3, 4, 5, 6, 8, 10, 12, 14, 16, and 17 have been KEGS focused fields, discovering to date 23 supernova, several of which been observed by SkyMapper (Rest et al. (in prep), Tucker et al. (in prep), Zenteno et al. (in prep)).

\begin{figure*}
\begin{center}
\resizebox{\textwidth}{!}{\includegraphics{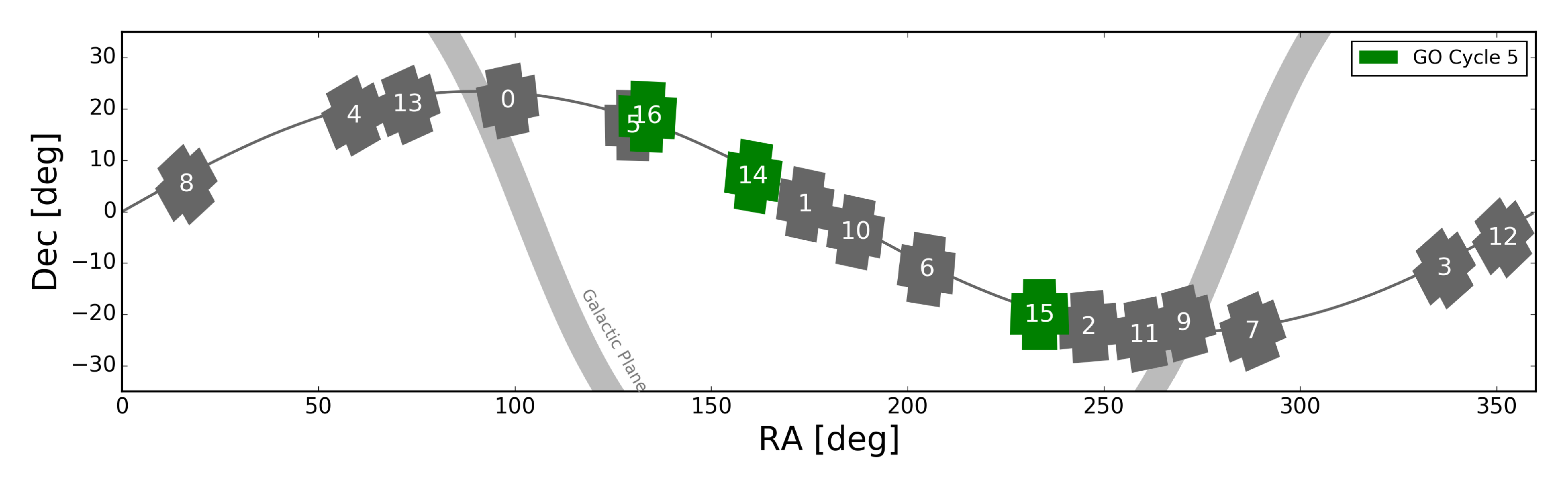}}
\end{center}
\caption{Figure courtesy of the NASA Kepler Guest Observer office.  These are the footprints for the K2 campaigns, which lie along the ecliptic, with the green fields being observed in 2017.  The Kepler Extra-Galactic Survey is monitoring, which SkyMapper is shadowing with ground-based multi-color observations, supernova in Campaigns 1, 3, 4, 5, 6, 8, 10, 12, 14, and 16. }
\label{fig:k2fields}
\end{figure*}


\section{SURVEY PERFORMANCE}

In this section we summarise the data taken for the SkyMapper
supernova survey, and evaluate its performance based on expectations from
detailed simulations of the survey history.


\subsection{Early survey} 

SkyMapper's performance has evolved over the commissioning period.
Early SkyMapper images were limited in quality by vibrations at $\sim 30$~Hz
driven by a resonance with the cooling system for the camera; these were
mitigated by modifications to change the telescope's resonant frequency.
During the Science Verification period for the SN Survey
(04 September 2013 to 09 March 2014), the median image quality was
$\sim$3.5~arcsec.  After March 2014, additional improvements to focus and
tracking resulted in a median image quality near 2~arcsec in \emph{gri} bands.
The readout overhead has also decreased, from a mean of 45 seconds during
Science Verification to 21 seconds as of April 2014.

From April 2014 through April 2015, the SkyMapper telescope was dedicated to
calibration of the SkyMapper standard star network (the Short Survey described in \S 2.
This work involved observations of bright stars, taking many repeat images of the sky with
short (5--10~sec) exposure times and no planned cadence.


\subsection{Image quality}

\begin{figure*}
\begin{center}
\resizebox{\textwidth}{!}{\includegraphics{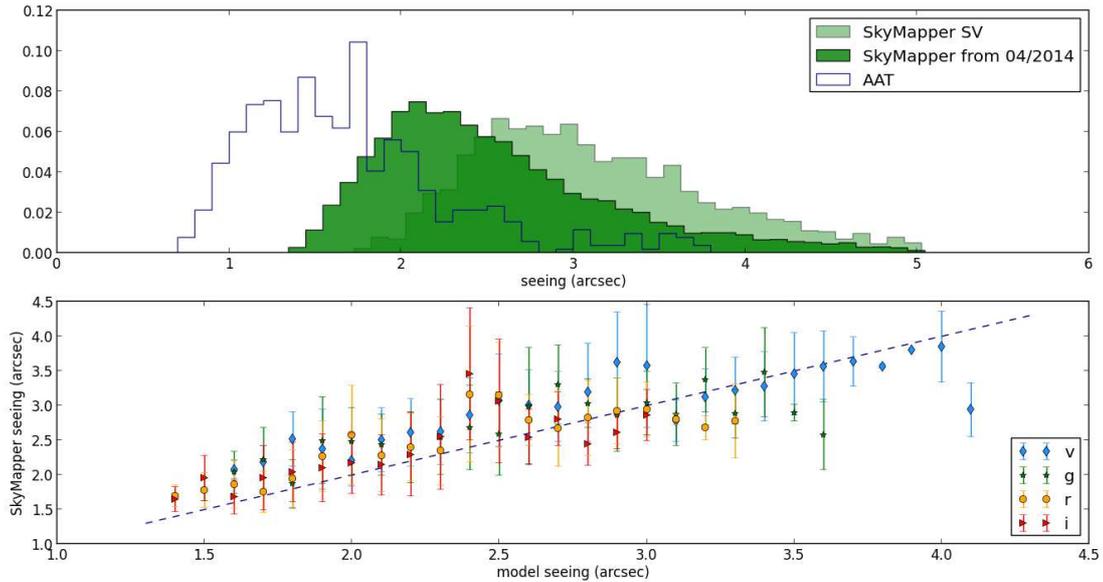}}
\end{center}
\caption{Top:  Histograms of \emph{g}-band seeing for the SkyMapper telescope
         during Science Verification (hatches) and after additional hardware
         intervention completed April 2014 (bold hatches), as compared to
         AAT seeing logs (open).  Bottom:  SkyMapper seeing in \emph{vgri} bands
         from April 2014 - May 2015, compared with predictions from the
         transfer function.}
\label{fig:seeing}
\end{figure*}

Figure~\ref{fig:seeing} shows the distribution of SkyMapper seeing in a
representative filter (\emph{g}), in comparison to the distribution obtained from
weather logs at the Anglo-Australian Telescope (AAT) assumed to be the natural
seeing of the Siding Spring site.  To assist us in running simulations to
determine the performance of our survey over a long historical period using
past weather logs, we developed a transfer function to predict SkyMapper
seeing from weather log entries.  The model takes into account nightly
variations in seeing (measured at the AAT), wavelength dependence,
and airmass:  for a filter with effective wavelength $\lambda$
and seeing $s_\lambda$,
\newcommand{\sskmz}{\ensuremath{s_{\mathrm{SM},0}}}
\begin{equation}
s_\lambda^2 = \sskmz^2 + \left[
        s_{AAT}^{1+\alpha}X^\gamma\left(\frac{\lambda}{5500\AA}
        \right)^\beta\right]^2,
\end{equation}
where $\sskmz = 1.1$~arcsec is a baseline seeing floor, and $\alpha$,
$\beta$, and $\gamma$ are coefficients characterizing atmospheric scattering.
For contemporary data (taken after April 2014), chi-square minimization
produces best-fit values $\alpha = -0.156$, $\beta = -0.5$, $\gamma = 0.8$. 

\subsection{Delays and limitations} 

\begin{figure}
\begin{center}
\resizebox{0.48\textwidth}{!}{\includegraphics{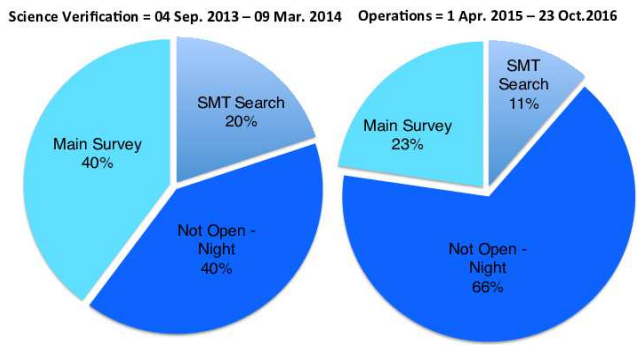}}
\end{center}
\caption{Impact of weather on Main Survey and Transient Survey operations
         as of 23 October 2016.}
\label{fig:weather}
\end{figure}

Through 2015 to early 2016 the SkyMapper telescope experienced various technical and software difficulties which delayed progress. These have typically halted survey operations for periods of 1-2 months at a time, during October-November 2015, and January and March 2016. Long periods of inclement weather (see Figure~\ref{fig:weather}) during the winter have frustrated the search for SNe, with both SkyMapper and the 2.3m telescope often being closed. This has adversely affected light curve quality and sampling, with numerous promising candidates fading before classification was possible, and/or left with large gaps in the light-curve. 

As more of the Main Survey footprint (Figure~\ref{fig:coverage}) is completed, more SMT fields will have deep references. This is decrease the time cost of building REF images during SMT time, compared to the early survey. The search has operated continuously from April 2016, during which supernovae have steadily been discovered and classified.


\section{FIRST RESULTS}

\subsection{First supernovae: Science Verification and Galaxy Zoo}

\begin{table*}
\center
\caption{Spectroscopically typed supernova discoveries during
         early SkyMapper operations}
\begin{tabular}{llrrlrl}
\hline 
Name & Disc. MJD (Phase)\nb{a} & RA & DEC & $z$ & Type & ATel \# \\
\hline 
\multicolumn{7}{c}{{\bf Science Verification (04 Sep 2013 -- 09 Mar 2014)}} \\
\hline 
SMT~J21413915$-$5643445 & 56573.6  ($-8$) & 21:41:39.15 & $-$56:43:44.5 & 0.142 &   Ia & 5480       \\
SMT~J23032187$-$6911189 & 56592.6  ($-3$) & 23:03:21.87 & $-$69:11:18.9 & 0.060 &   Ia & 5521\nb{b} \\
SMT~J03054854$-$2850370 & 56626.6         & 03:05:48.54 & $-$28:50:37.0 & 0.050 &  IIn & 5622 \\
SMT~J03253351$-$5344190 & 56627.6 ($-12$) & 03:25:33.51 & $-$53:44:19.0 & 0.055 &   Ia & 5641       \\
SMT~J00570507$-$3626231 & 56628.8  ($+3$) & 00:57:05.07 & $-$36:26:23.1 & 0.057 &   Ia & 5620       \\
SMT~J03264288$-$3438055 & 56638.6 ($+14)$ & 03:26:42.88 & $-$34:38:05.5 & 0.1~~ &   Ia & 5602\nb{c} \\
SMT~J03101002$-$3637448 & 56653.6 ($+11)$ & 03:10:10.02 & $-$36:37:44.8 & 0.070 &   Ia & 5650\nb{d} \\
SN~2013hx               & 56653.8         & 01:35:32.83 & $-$57:57:50.6 & 0.130 & SLSN & 5912 \\
SMT~J04043173$-$6350154 & 56664.5  ($+2)$ & 04:04:31.73 & $-$63:50:15.4 & 0.1~~ &   Ia & 5748\nb{e} \\
SMT~J05451320$-$4735425 & 56666.7  ($-5)$ & 05:45:13.20 & $-$47:35:42.5 & 0.050 &   Ia &  ---       \\
\hline 
\multicolumn{7}{c}{{\bf Zooniverse Campaign (12--22 Mar 2015)}} \\
\hline 
SMT~J10310056$-$3658262 & 57094.5  ($+0$) & 10:31:00.56 & $-$36:58:26.2 & 0.035 &   Ia & 7261 \\
SMT~J13254308$-$2932269 & 57094.6         & 13:25:43.08 & $-$29:32:26.9 & 0.040 &   Ic & 7254 \\
SMT~J13545988$-$2820020 & 57094.6  ($+0$) & 13:54:59.88 & $-$28:20:02.0 & 0.038 &   Ia & 7261 \\
SMT~J14323134$-$1339275 & 57095.7         & 14:32:31.34 & $-$13:39:27.5 & 0.021 &  IIb & 7261\nb{f} \\
SMT~J13481313$-$3325189 & 57094.6 ($+21$) & 13:48:13.13 & $-$33:25:18.9 &       &   Ia &  ---\nb{g} \\
\hline 
\end{tabular}
\medskip \\
\flushleft
$^a$ Phase in days relative to $B$-band maximum light (type Ia only). \\
$^b$ Discovered independently and first confirmed as PSNJ23032177$-$6911185 by the CHASE survey. \\
$^c$ Discovered independently and first confirmed as LSQ13dby. \\
$^d$ Discovered independently and first confirmed as LSQ13dkp. \\
$^e$ Discovered independently as OGLE-2014-SN-002. \\
$^f$ Discovered independently as LSQ15rw. \\
$^g$ Classified by LCOGT as a Ia well after maximum light (G.~Hosseinzadeh, priv. comm). \\

\label{tbl:sne}
\end{table*}

\begin{figure*}
\begin{center}
\resizebox{0.8\textwidth}{!}{\includegraphics{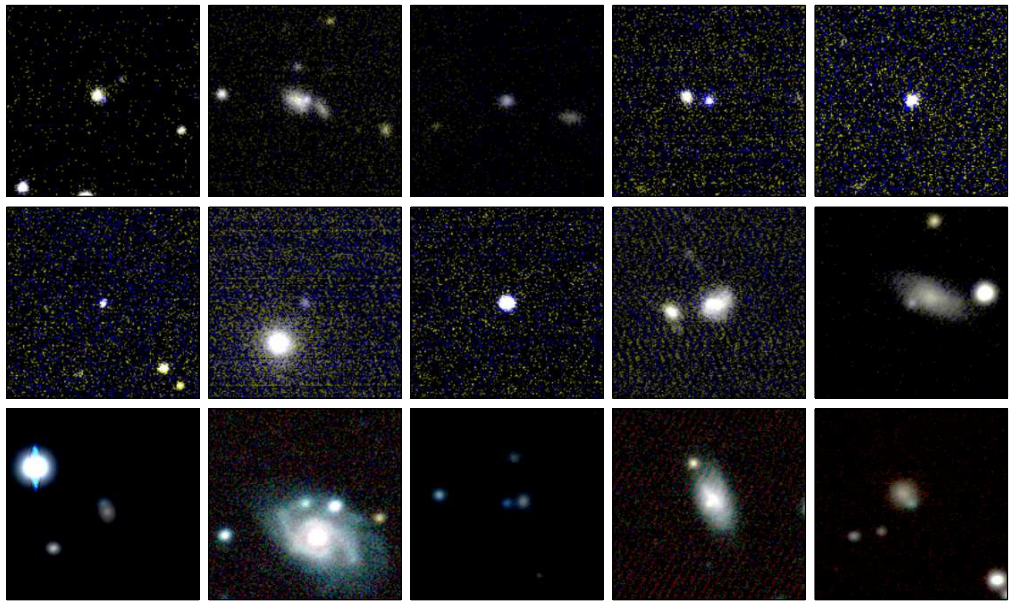}}
\end{center}
\caption{Color composite thumbnail images of a selection of early  SkyMapper supernova discoveries.}
\label{fig:thumbs}
\end{figure*}

Table~\ref{tbl:sne} reports the types and numbers of confirmed SNe discovered
(see Figure~\ref{fig:thumbs} for color composite images) during early SkyMapper
operations.  During Science Verification, the Search produced 10 spectroscopically
confirmed supernovae, among them 8 SNe~Ia.

\begin{figure}
\begin{center}
\resizebox{0.5\textwidth}{!}{\includegraphics{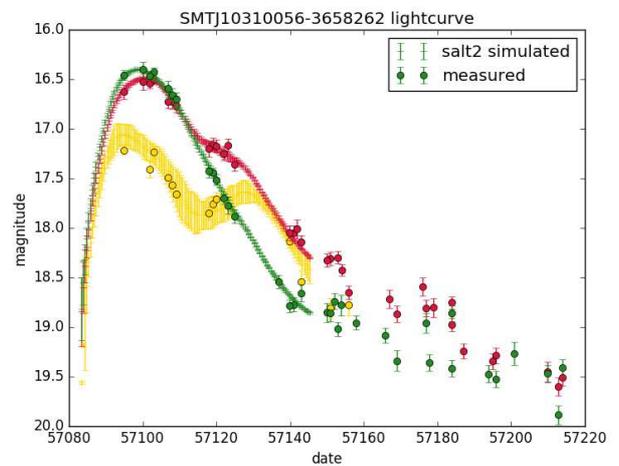}}
\end{center}
\caption{Lightcurve in \emph{gri} colours of SMTJ10310056$-$3658262 discovered during the Galaxy Zoo campaign, shown with SALT2 fit.}
\label{fig:Ia}
\end{figure}

Supernova observations resumed during the period 12--20 March 2015,
during which SkyMapper performed an intensive observing campaign with a short
cadence of 1--2 days, as an outreach effort in partnership with the
Galaxy Zoo citizen science community.  An additional five spectroscopically
confirmed SNe were discovered in this campaign, including three SNe~Ia.
Thumbnails of candidates found in the
search appeared on the Galaxy Zoo website, and were processed by volunteers
using the decision tree described in \citet{snzoo}.  Popular candidates were
inspected by the authors and submitted in real time for spectroscopic
classification by the PESSTO collaboration using the ESO NTT 3.6-m telescope
at La Silla. An example of typical light curve quality in the early survey is shown in Figure~\ref{fig:Ia} for SMTJ10310056$-$3658262, a type Ia supernova discovered during the Galaxy Zoo campaign.

\begin{figure*}
\begin{center}
\resizebox{1.0\textwidth}{!}{\includegraphics{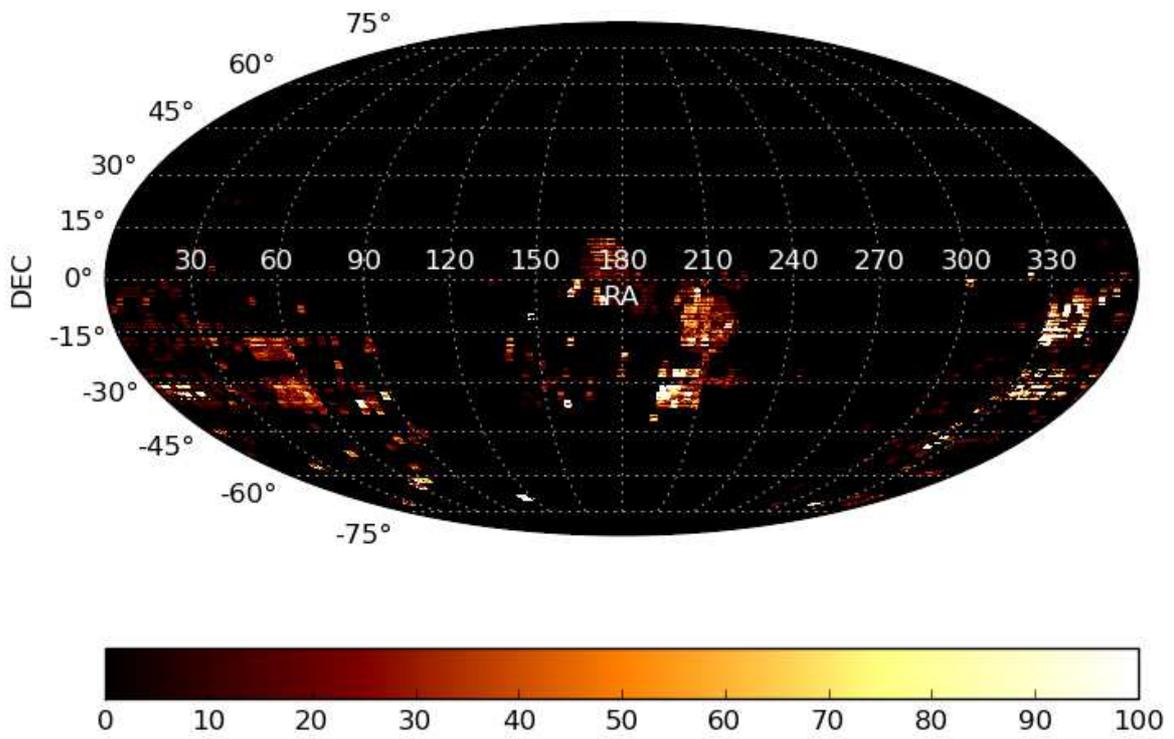}}
\end{center}
\caption{Map of cumulative sky coverage for the SkyMapper supernova search
         as of 20 October 2016.}
\label{fig:coverage}
\end{figure*}

Figure~\ref{fig:coverage} shows a map of the total supernova survey coverage
to date.  A total of 393 SkyMapper fields, or 2250~\sqdeg, have been
observed, with a mean of 40 visits since the beginning of Science Verification.
The expected number $N$ of supernovae to be found in the survey can be
estimated by Monte Carlo, integrating
\begin{equation}
N = \int d\Omega \int dt \int dz \,
   \frac{\partial V}{\partial z} \eta(z, \Omega, t)
\end{equation}
over time $t$, the survey footprint $\Omega$, and redshift $z$; here
$dV$ is the co-moving volume element at redshift $z$, and $\eta$ is the survey
efficiency for discovery of supernovae including the particulars of the timing
and depth of each image relative to a randomly generated set of supernova
light curves.  We estimate an effective significance threshold of 9$\sigma$ for single-epoch detections based on the empirical Real/Bogus classifier efficiency curve (Figure~\ref{fig:rb-eff}).
We use the historical cadences and upper limits from successful subtractions.

Based on these assumptions, in the period April--November 2016, when the instrument configuration was stable and the survey was working well, we expect a total of
$65 \pm 8$ SNe~Ia to be found by the survey in that period, assuming 2 significant
detections on separate nights in any filter were necessary for detection.
If instead simultaneous detections in both $g$ and $r$ are required, we expect
$41 \pm 6$ SNe~Ia instead.  Our total of 13 spectroscopically confirmed SNe~Ia during this period is lower than expected, with a number of possible factors contributing:
losses in the classifier at the pixel level (not simulated at this stage),
departures of the SkyMapper PSF from the assumed (round Gaussian) PSF up through
Science Verification, edge effects due to slight pointing differences between
NEW and REF, and spectroscopic selection.  Of these effects, we expect that
selection for spectroscopic follow-up is probably the largest of these effects,
due to weather-induced gaps in light curve coverage and follow-up availability.
We identify an additional 39 transient objects found in our survey in 2016 with appropriate light curve timescales and visible host galaxies, but no spectra.

\subsection{Unique peculiar objects}
\label{sec:weirdo}

The SkyMapper Transient Search is also discovering other types of supernova and stellar transients, such as superluminous supernovae \citep{quimby11,galyam12}
The first notable SkyMapper object was indeed one of these, SN~2013hx (discovered as SMT~J013533283$-$5757506), a superluminous supernova initially similar to SN~2010gx \citep{pastorello10} but displaying broad H$\alpha$ emission at late times
\citep{Inserra16}. SN~2013hx was discovered by SkyMapper at MJD~56657.6 (2013~Dec~31~UT), and reached reached peak magnitudes $g = 16.9$, $r = 17.0$ at MJD~56683.5 
(2014~Jan~26~UT). At a redshift of 0.130, SN~2013hx was the closest superluminous type II
supernova discovered to date \citep{nicholl14}, presenting an excellent opportunity for late-time observations. SN~2013hx was also included in a recent study of superluminous supernova light curves \citep{nicholl15}.

We are also sensitive to more exotic transients, such as faint calcium-rich transients \citep{kasliwal12}, which occur preferentially in low-surface-brightness, star-forming host galaxies \citep{neill11} or on the outskirts of larger galaxies not monitored by targeted searches. \citep{yuan13} We have already discovered two exotic transients: SN 2015J, a possible SN impostor or magnetar-powered SN Ic (Tucker et al. in prep), and an object similar to \cite{arcavi16} in the so-called superluminous-gap (Zhang et al. in prep). We may also be sensitive to shock interaction with SN Ia companions, as observed in SN~2016hhd. (M{\"o}ller et al. in prep)

Moreover, we are also triggering on other exotic multi-wavelength events such as Fast Radio Bursts (FRBs) and Gravitational Wave (GW) alerts. Additionally, the data should also be useful for studying variable stars, active galactic nuclei, and other types of transients, and can be co-added to provide deep exposures of the search area.

 \section{SUMMARY} 

This work presents the SkyMapper Transient Survey, including the software specific to the Transient Search and the planned survey strategy. The former involves a sophisticated image subtraction pipeline with a machine learning classifier and a web admin interface for human input. We describe the early performance of the Survey,
which is steadily assembling a small sample of SNe~Ia for inclusion in a low-redshift cosmology sample (to be detailed in a forthcoming data release paper), and in addition discovering peculiar objects interesting in their own right (Tucker et al., Zhang et al., M{\"o}ller et al.; all in preparation).

\begin{acknowledgements}

Parts of this research were conducted by the Australian Research Council
Centre of Excellence for All-sky Astrophysics (CAASTRO), through project
number CE110001020. RS acknowledges support from ARC Laureate Grant FL0992131.
This research was made possible through the use of the AAVSO Photometric
All-sky Survey (APASS), funded by the Robert Martin Ayers Sciences Fund.
We gratefully acknowledge the assistance of over 40,000 online volunteers
for the Snapshot Supernova program in March 2015, run in partnership with
the Zooniverse citizen science portal.  We also thank Andy Howell and the
LCOGT team for providing us with a spectroscopic classification for
SMT~J13481313$-$3325189.  We acknowledge and thank PhD students: F. Panther,  R. Ridden-Harper and  N.E. Sommer as well contributions by summer students: P. Armstrong, G. Taylor, E. Moore, C. Bray and Y. Chen.

\end{acknowledgements}





\bibliographystyle{apj}
\bibliography{subpipe_pasa}

\begin{thebibliography}{}
\expandafter\ifx\csname natexlab\endcsname\relax\def\natexlab#1{#1}\fi

\bibitem[{{Arcavi} {et~al.}(2016){Arcavi}, {Wolf}, {Howell}, {Bildsten},
  {Leloudas}, {Hardin}, {Prajs}, {Perley}, {Svirski}, {Gal-Yam}, {Katz},
  {McCully}, {Cenko}, {Lidman}, {Sullivan}, {Valenti}, {Astier}, {Balland},
  {Carlberg}, {Conley}, {Fouchez}, {Guy}, {Pain}, {Palanque-Delabrouille},
  {Perrett}, {Pritchet}, {Regnault}, {Rich}, \& {Ruhlmann-Kleider}}]{arcavi16}
{Arcavi}, I., {Wolf}, W.~M., {Howell}, D.~A., {et~al.} 2016, \apj, 819, 35

\bibitem[{{Baltay} {et~al.}(2013){Baltay}, {Rabinowitz}, {Hadjiyska}, {Walker},
  {Nugent}, {Coppi}, {Ellman}, {Feindt}, {McKinnon}, {Horowitz}, \&
  {Effron}}]{lsq}
{Baltay}, C., {Rabinowitz}, D., {Hadjiyska}, E., {et~al.} 2013, \pasp, 125, 683

\bibitem[{{Berthier} {et~al.}(2006){Berthier}, {Vachier}, {Thuillot},
  {Fernique}, {Ochsenbein}, {Genova}, {Lainey}, \& {Arlot}}]{skybot}
{Berthier}, J., {Vachier}, F., {Thuillot}, W., {et~al.} 2006, in Astronomical
  Society of the Pacific Conference Series, Vol. 351, Astronomical Data
  Analysis Software and Systems XV, ed. C.~{Gabriel}, C.~{Arviset}, D.~{Ponz},
  \& S.~{Enrique}, 367--+

\bibitem[{{Bertin} \& {Arnouts}(1996)}]{sex}
{Bertin}, E., \& {Arnouts}, S. 1996, \aaps, 117, 393

\bibitem[{{Bertin} {et~al.}(2002){Bertin}, {Mellier}, {Radovich}, {Missonnier},
  {Didelon}, \& {Morin}}]{swarp}
{Bertin}, E., {Mellier}, Y., {Radovich}, M., {et~al.} 2002, in Astronomical
  Society of the Pacific Conference Series, Vol. 281, Astronomical Data
  Analysis Software and Systems XI, ed. D.~A. {Bohlender}, D.~{Durand}, \&
  T.~H. {Handley}, 228

\bibitem[{{Bessell} {et~al.}(2011){Bessell}, {Bloxham}, {Schmidt}, {Keller},
  {Tisserand}, \& {Francis}}]{bessell11}
{Bessell}, M., {Bloxham}, G., {Schmidt}, B., {et~al.} 2011, \pasp, 123, 789

\bibitem[{{Betoule} {et~al.}(2013){Betoule}, {Marriner}, {Regnault},
  {Cuillandre}, {Astier}, {Guy}, {Balland}, {El Hage}, {Hardin}, {Kessler}, {Le
  Guillou}, {Mosher}, {Pain}, {Rocci}, {Sako}, \& {Schahmaneche}}]{betoule13}
{Betoule}, M., {Marriner}, J., {Regnault}, N., {et~al.} 2013, \aap, 552, A124

\bibitem[{{Betoule} {et~al.}(2014){Betoule}, {Kessler}, {Guy}, {Mosher},
  {Hardin}, {Biswas}, {Astier}, {El-Hage}, {Konig}, {Kuhlmann}, {Marriner},
  {Pain}, {Regnault}, {Balland}, {Bassett}, {Brown}, {Campbell}, {Carlberg},
  {Cellier-Holzem}, {Cinabro}, {Conley}, {D'Andrea}, {DePoy}, {Doi}, {Ellis},
  {Fabbro}, {Filippenko}, {Foley}, {Frieman}, {Fouchez}, {Galbany}, {Goobar},
  {Gupta}, {Hill}, {Hlozek}, {Hogan}, {Hook}, {Howell}, {Jha}, {Le Guillou},
  {Leloudas}, {Lidman}, {Marshall}, {M{\"o}ller}, {Mour{\~a}o}, {Neveu},
  {Nichol}, {Olmstead}, {Palanque-Delabrouille}, {Perlmutter}, {Prieto},
  {Pritchet}, {Richmond}, {Riess}, {Ruhlmann-Kleider}, {Sako}, {Schahmaneche},
  {Schneider}, {Smith}, {Sollerman}, {Sullivan}, {Walton}, \&
  {Wheeler}}]{betoule14}
{Betoule}, M., {Kessler}, R., {Guy}, J., {et~al.} 2014, \aap, 568, A22

\bibitem[{{Bloom} {et~al.}(2012){Bloom}, {Richards}, {Nugent}, {Quimby},
  {Kasliwal}, {Starr}, {Poznanski}, {Ofek}, {Cenko}, {Butler}, {Kulkarni},
  {Gal-Yam}, \& {Law}}]{bloom12}
{Bloom}, J.~S., {Richards}, J.~W., {Nugent}, P.~E., {et~al.} 2012, \pasp, 124,
  1175

\bibitem[{{Brown} {et~al.}(2013){Brown}, {Baliber}, {Bianco}, {Bowman},
  {Burleson}, {Conway}, {Crellin}, {Depagne}, {De Vera}, {Dilday}, {Dragomir},
  {Dubberley}, {Eastman}, {Elphick}, {Falarski}, {Foale}, {Ford}, {Fulton},
  {Garza}, {Gomez}, {Graham}, {Greene}, {Haldeman}, {Hawkins}, {Haworth},
  {Haynes}, {Hidas}, {Hjelstrom}, {Howell}, {Hygelund}, {Lister}, {Lobdill},
  {Martinez}, {Mullins}, {Norbury}, {Parrent}, {Paulson}, {Petry}, {Pickles},
  {Posner}, {Rosing}, {Ross}, {Sand}, {Saunders}, {Shobbrook}, {Shporer},
  {Street}, {Thomas}, {Tsapras}, {Tufts}, {Valenti}, {Vander Horst}, {Walker},
  {White}, \& {Willis}}]{lcogt}
{Brown}, T.~M., {Baliber}, N., {Bianco}, F.~B., {et~al.} 2013, \pasp, 125, 1031

\bibitem[{{Burns} {et~al.}(2014){Burns}, {Stritzinger}, {Phillips}, {Hsiao},
  {Contreras}, {Persson}, {Folatelli}, {Boldt}, {Campillay}, {Castell{\'o}n},
  {Freedman}, {Madore}, {Morrell}, {Salgado}, \& {Suntzeff}}]{burns14}
{Burns}, C.~R., {Stritzinger}, M., {Phillips}, M.~M., {et~al.} 2014, \apj, 789,
  32

\bibitem[{{Childress} {et~al.}(2013){Childress}, {Aldering}, {Antilogus},
  {Aragon}, {Bailey}, {Baltay}, {Bongard}, {Buton}, {Canto}, {Cellier-Holzem},
  {Chotard}, {Copin}, {Fakhouri}, {Gangler}, {Guy}, {Hsiao}, {Kerschhaggl},
  {Kim}, {Kowalski}, {Loken}, {Nugent}, {Paech}, {Pain}, {Pecontal}, {Pereira},
  {Perlmutter}, {Rabinowitz}, {Rigault}, {Runge}, {Scalzo}, {Smadja}, {Tao},
  {Thomas}, {Weaver}, \& {Wu}}]{childress13}
{Childress}, M., {Aldering}, G., {Antilogus}, P., {et~al.} 2013, \apj, 770, 108

\bibitem[{{Conley} {et~al.}(2011){Conley}, {Guy}, {Sullivan}, {Regnault},
  {Astier}, {Balland}, {Basa}, {Carlberg}, {Fouchez}, {Hardin}, {Hook},
  {Howell}, {Pain}, {Palanque-Delabrouille}, {Perrett}, {Pritchet}, {Rich},
  {Ruhlmann-Kleider}, {Balam}, {Baumont}, {Ellis}, {Fabbro}, {Fakhouri},
  {Fourmanoit}, {Gonz{\'a}lez-Gait{\'a}n}, {Graham}, {Hudson}, {Hsiao},
  {Kronborg}, {Lidman}, {Mourao}, {Neill}, {Perlmutter}, {Ripoche}, {Suzuki},
  \& {Walker}}]{conley11}
{Conley}, A., {Guy}, J., {Sullivan}, M., {et~al.} 2011, \apjs, 192, 1

\bibitem[{{Contreras} {et~al.}(2010){Contreras}, {Hamuy}, {Phillips},
  {Folatelli}, {Suntzeff}, {Persson}, {Stritzinger}, {Boldt}, {Gonz{\'a}lez},
  {Krzeminski}, {Morrell}, {Roth}, {Salgado}, {Jos{\'e} Maureira}, {Burns},
  {Freedman}, {Madore}, {Murphy}, {Wyatt}, {Li}, \& {Filippenko}}]{csp}
{Contreras}, C., {Hamuy}, M., {Phillips}, M.~M., {et~al.} 2010, \aj, 139, 519

\bibitem[{{Dark Energy Survey Collaboration} {et~al.}(2016){Dark Energy Survey
  Collaboration}, {Abbott}, {Abdalla}, {Aleksi{\'c}}, {Allam}, {Amara},
  {Bacon}, {Balbinot}, {Banerji}, {Bechtol}, {Benoit-L{\'e}vy}, {Bernstein},
  {Bertin}, {Blazek}, {Bonnett}, {Bridle}, {Brooks}, {Brunner}, {Buckley-Geer},
  {Burke}, {Caminha}, {Capozzi}, {Carlsen}, {Carnero-Rosell}, {Carollo},
  {Carrasco-Kind}, {Carretero}, {Castander}, {Clerkin}, {Collett}, {Conselice},
  {Crocce}, {Cunha}, {D'Andrea}, {da Costa}, {Davis}, {Desai}, {Diehl},
  {Dietrich}, {Dodelson}, {Doel}, {Drlica-Wagner}, {Estrada}, {Etherington},
  {Evrard}, {Fabbri}, {Finley}, {Flaugher}, {Foley}, {Fosalba}, {Frieman},
  {Garc{\'{\i}}a-Bellido}, {Gaztanaga}, {Gerdes}, {Giannantonio}, {Goldstein},
  {Gruen}, {Gruendl}, {Guarnieri}, {Gutierrez}, {Hartley}, {Honscheid}, {Jain},
  {James}, {Jeltema}, {Jouvel}, {Kessler}, {King}, {Kirk}, {Kron}, {Kuehn},
  {Kuropatkin}, {Lahav}, {Li}, {Lima}, {Lin}, {Maia}, {Makler}, {Manera},
  {Maraston}, {Marshall}, {Martini}, {McMahon}, {Melchior}, {Merson}, {Miller},
  {Miquel}, {Mohr}, {Morice-Atkinson}, {Naidoo}, {Neilsen}, {Nichol}, {Nord},
  {Ogando}, {Ostrovski}, {Palmese}, {Papadopoulos}, {Peiris}, {Peoples},
  {Percival}, {Plazas}, {Reed}, {Refregier}, {Romer}, {Roodman}, {Ross},
  {Rozo}, {Rykoff}, {Sadeh}, {Sako}, {S{\'a}nchez}, {Sanchez}, {Santiago},
  {Scarpine}, {Schubnell}, {Sevilla-Noarbe}, {Sheldon}, {Smith}, {Smith},
  {Soares-Santos}, {Sobreira}, {Soumagnac}, {Suchyta}, {Sullivan}, {Swanson},
  {Tarle}, {Thaler}, {Thomas}, {Thomas}, {Tucker}, {Vieira}, {Vikram},
  {Walker}, {Wechsler}, {Weller}, {Wester}, {Whiteway}, {Wilcox}, {Yanny},
  {Zhang}, \& {Zuntz}}]{des}
{Dark Energy Survey Collaboration}, {Abbott}, T., {Abdalla}, F.~B., {et~al.}
  2016, \mnras, 460, 1270

\bibitem[{{Davis} {et~al.}(2011){Davis}, {Hui}, {Frieman}, {Haugb{\o}lle},
  {Kessler}, {Sinclair}, {Sollerman}, {Bassett}, {Marriner}, {M{\"o}rtsell},
  {Nichol}, {Richmond}, {Sako}, {Schneider}, \& {Smith}}]{davis11}
{Davis}, T.~M., {Hui}, L., {Frieman}, J.~A., {et~al.} 2011, \apj, 741, 67

\bibitem[{{Djorgovski} {et~al.}(2008){Djorgovski}, {Baltay}, {Mahabal},
  {Drake}, {Williams}, {Rabinowitz}, {Graham}, {Donalek}, {Glikman}, {Bauer},
  {Scalzo}, \& {Ellman}}]{pq}
{Djorgovski}, S.~G., {Baltay}, C., {Mahabal}, A.~A., {et~al.} 2008,
  Astronomische Nachrichten, 329, 263

\bibitem[{{Dopita} {et~al.}(2007){Dopita}, {Hart}, {McGregor}, {Oates},
  {Bloxham}, \& {Jones}}]{wifes}
{Dopita}, M., {Hart}, J., {McGregor}, P., {et~al.} 2007, \apss, 310, 255

\bibitem[{{Drake} {et~al.}(2009){Drake}, {Djorgovski}, {Mahabal}, {Beshore},
  {Larson}, {Graham}, {Williams}, {Christensen}, {Catelan}, {Boattini},
  {Gibbs}, {Hill}, \& {Kowalski}}]{crts}
{Drake}, A.~J., {Djorgovski}, S.~G., {Mahabal}, A., {et~al.} 2009, \apj, 696,
  870

\bibitem[{{Filippenko} {et~al.}(2001){Filippenko}, {Li}, {Treffers}, \&
  {Modjaz}}]{loss2}
{Filippenko}, A.~V., {Li}, W.~D., {Treffers}, R.~R., \& {Modjaz}, M. 2001, in
  Astronomical Society of the Pacific Conference Series, Vol. 246, IAU Colloq.
  183: Small Telescope Astronomy on Global Scales, ed. B.~{Paczynski}, W.-P.
  {Chen}, \& C.~{Lemme}, 121

\bibitem[{{Gal-Yam}(2012)}]{galyam12}
{Gal-Yam}, A. 2012, Science, 337, 927

\bibitem[{{Haugb{\o}lle} {et~al.}(2007){Haugb{\o}lle}, {Hannestad}, {Thomsen},
  {Fynbo}, {Sollerman}, \& {Jha}}]{haugbolle07}
{Haugb{\o}lle}, T., {Hannestad}, S., {Thomsen}, B., {et~al.} 2007, \apj, 661,
  650

\bibitem[{{Henden} \& {Munari}(2014)}]{apass}
{Henden}, A., \& {Munari}, U. 2014, Contributions of the Astronomical
  Observatory Skalnate Pleso, 43, 518

\bibitem[{{Hicken} {et~al.}(2009){Hicken}, {Challis}, {Jha}, {Kirshner},
  {Matheson}, {Modjaz}, {Rest}, {Wood-Vasey}, {Bakos}, {Barton}, {Berlind},
  {Bragg}, {Brice{\~n}o}, {Brown}, {Caldwell}, {Calkins}, {Cho}, {Ciupik},
  {Contreras}, {Dendy}, {Dosaj}, {Durham}, {Eriksen}, {Esquerdo}, {Everett},
  {Falco}, {Fernandez}, {Gaba}, {Garnavich}, {Graves}, {Green}, {Groner},
  {Hergenrother}, {Holman}, {Hradecky}, {Huchra}, {Hutchison}, {Jerius},
  {Jordan}, {Kilgard}, {Krauss}, {Luhman}, {Macri}, {Marrone}, {McDowell},
  {McIntosh}, {McNamara}, {Megeath}, {Mochejska}, {Munoz}, {Muzerolle},
  {Naranjo}, {Narayan}, {Pahre}, {Peters}, {Peterson}, {Rines}, {Ripman},
  {Roussanova}, {Schild}, {Sicilia-Aguilar}, {Sokoloski}, {Smalley}, {Smith},
  {Spahr}, {Stanek}, {Barmby}, {Blondin}, {Stubbs}, {Szentgyorgyi}, {Torres},
  {Vaz}, {Vikhlinin}, {Wang}, {Westover}, {Woods}, \& {Zhao}}]{cfa3}
{Hicken}, M., {Challis}, P., {Jha}, S., {et~al.} 2009, \apj, 700, 331

\bibitem[{{Hicken} {et~al.}(2012){Hicken}, {Challis}, {Kirshner}, {Rest},
  {Cramer}, {Wood-Vasey}, {Bakos}, {Berlind}, {Brown}, {Caldwell}, {Calkins},
  {Currie}, {de Kleer}, {Esquerdo}, {Everett}, {Falco}, {Fernandez},
  {Friedman}, {Groner}, {Hartman}, {Holman}, {Hutchins}, {Keys}, {Kipping},
  {Latham}, {Marion}, {Narayan}, {Pahre}, {Pal}, {Peters}, {Perumpilly},
  {Ripman}, {Sipocz}, {Szentgyorgyi}, {Tang}, {Torres}, {Vaz}, {Wolk}, \&
  {Zezas}}]{cfa4}
{Hicken}, M., {Challis}, P., {Kirshner}, R.~P., {et~al.} 2012, \apjs, 200, 12

\bibitem[{{Howell} {et~al.}(2007){Howell}, {Sullivan}, {Conley}, \&
  {Carlberg}}]{howell07}
{Howell}, D.~A., {Sullivan}, M., {Conley}, A., \& {Carlberg}, R. 2007, \apjl,
  667, L37

\bibitem[{{Hui} \& {Greene}(2006)}]{hg06}
{Hui}, L., \& {Greene}, P.~B. 2006, \prd, 73, 123526

\bibitem[{{Inserra} {et~al.}(2016){Inserra}, {Smartt}, {Gall}, {Leloudas},
  {Chen}, {Schulze}, {Jerkstarnd}, {Nicholl}, {Anderson}, {Arcavi}, {Benetti},
  {Cartier}, {Childress}, {Della Valle}, {Flewelling}, {Fraser}, {Gal-Yam},
  {Gutierrez}, {Hosseinzadeh}, {Howell}, {Huber}, {Kankare}, {Magnier},
  {Maguire}, {McCully}, {Prajs}, {Primak}, {Scalzo}, {Schmidt}, {Smith},
  {Tucker}, {Valenti}, {Wilman}, {Young}, \& {Yuan}}]{Inserra16}
{Inserra}, C., {Smartt}, S.~J., {Gall}, E.~E.~E., {et~al.} 2016, ArXiv
  e-prints, arXiv:1604.01226

\bibitem[{{Jha} {et~al.}(2006){Jha}, {Kirshner}, {Challis}, {Garnavich},
  {Matheson}, {Soderberg}, {Graves}, {Hicken}, {Alves}, {Arce}, {Balog},
  {Barmby}, {Barton}, {Berlind}, {Bragg}, {Brice{\~n}o}, {Brown}, {Buckley},
  {Caldwell}, {Calkins}, {Carter}, {Concannon}, {Donnelly}, {Eriksen},
  {Fabricant}, {Falco}, {Fiore}, {Garcia}, {G{\'o}mez}, {Grogin}, {Groner},
  {Groot}, {Haisch}, {Hartmann}, {Hergenrother}, {Holman}, {Huchra},
  {Jayawardhana}, {Jerius}, {Kannappan}, {Kim}, {Kleyna}, {Kochanek},
  {Koranyi}, {Krockenberger}, {Lada}, {Luhman}, {Luu}, {Macri}, {Mader},
  {Mahdavi}, {Marengo}, {Marsden}, {McLeod}, {McNamara}, {Megeath}, {Moraru},
  {Mossman}, {Muench}, {Mu{\~n}oz}, {Muzerolle}, {Naranjo}, {Nelson-Patel},
  {Pahre}, {Patten}, {Peters}, {Peters}, {Raymond}, {Rines}, {Schild},
  {Sobczak}, {Spahr}, {Stauffer}, {Stefanik}, {Szentgyorgyi}, {Tollestrup},
  {V{\"a}is{\"a}nen}, {Vikhlinin}, {Wang}, {Willner}, {Wolk}, {Zajac}, {Zhao},
  \& {Stanek}}]{cfa2}
{Jha}, S., {Kirshner}, R.~P., {Challis}, P., {et~al.} 2006, \aj, 131, 527

\bibitem[{{Kaiser} {et~al.}(2010){Kaiser}, {Burgett}, {Chambers}, {Denneau},
  {Heasley}, {Jedicke}, {Magnier}, {Morgan}, {Onaka}, \& {Tonry}}]{ps1}
{Kaiser}, N., {Burgett}, W., {Chambers}, K., {et~al.} 2010, in Society of
  Photo-Optical Instrumentation Engineers (SPIE) Conference Series, Vol. 7733,
  Society of Photo-Optical Instrumentation Engineers (SPIE) Conference Series,
  0

\bibitem[{{Kasliwal} {et~al.}(2012){Kasliwal}, {Kulkarni}, {Gal-Yam}, {Nugent},
  {Sullivan}, {Bildsten}, {Yaron}, {Perets}, {Arcavi}, {Ben-Ami}, {Bhalerao},
  {Bloom}, {Cenko}, {Filippenko}, {Frail}, {Ganeshalingam}, {Horesh}, {Howell},
  {Law}, {Leonard}, {Li}, {Ofek}, {Polishook}, {Poznanski}, {Quimby},
  {Silverman}, {Sternberg}, \& {Xu}}]{kasliwal12}
{Kasliwal}, M.~M., {Kulkarni}, S.~R., {Gal-Yam}, A., {et~al.} 2012, \apj, 755,
  161

\bibitem[{{Keller} {et~al.}(2007){Keller}, {Schmidt}, {Bessell}, {Conroy},
  {Francis}, {Granlund}, {Kowald}, {Oates}, {Martin-Jones}, {Preston},
  {Tisserand}, {Vaccarella}, \& {Waterson}}]{keller07}
{Keller}, S.~C., {Schmidt}, B.~P., {Bessell}, M.~S., {et~al.} 2007, \pasa, 24,
  1

\bibitem[{{Keller} {et~al.}(2014){Keller}, {Bessell}, {Frebel}, {Casey},
  {Asplund}, {Jacobson}, {Lind}, {Norris}, {Yong}, {Heger}, {Magic}, {da
  Costa}, {Schmidt}, \& {Tisserand}}]{keller14}
{Keller}, S.~C., {Bessell}, M.~S., {Frebel}, A., {et~al.} 2014, \nat, 506, 463

\bibitem[{{Kelly} {et~al.}(2015){Kelly}, {Filippenko}, {Burke}, {Hicken},
  {Ganeshalingam}, \& {Zheng}}]{kelly15}
{Kelly}, P.~L., {Filippenko}, A.~V., {Burke}, D.~L., {et~al.} 2015, Science,
  347, 1459

\bibitem[{{Kelly} {et~al.}(2010){Kelly}, {Hicken}, {Burke}, {Mandel}, \&
  {Kirshner}}]{kelly10}
{Kelly}, P.~L., {Hicken}, M., {Burke}, D.~L., {Mandel}, K.~S., \& {Kirshner},
  R.~P. 2010, \apj, 715, 743

\bibitem[{{Kim} {et~al.}(2004){Kim}, {Linder}, {Miquel}, \& {Mostek}}]{kim04}
{Kim}, A.~G., {Linder}, E.~V., {Miquel}, R., \& {Mostek}, N. 2004, \mnras, 347,
  909

\bibitem[{{Law} {et~al.}(2009){Law}, {Kulkarni}, {Dekany}, {Ofek}, {Quimby},
  {Nugent}, {Surace}, {Grillmair}, {Bloom}, {Kasliwal}, {Bildsten}, {Brown},
  {Cenko}, {Ciardi}, {Croner}, {Djorgovski}, {van Eyken}, {Filippenko}, {Fox},
  {Gal-Yam}, {Hale}, {Hamam}, {Helou}, {Henning}, {Howell}, {Jacobsen},
  {Laher}, {Mattingly}, {McKenna}, {Pickles}, {Poznanski}, {Rahmer}, {Rau},
  {Rosing}, {Shara}, {Smith}, {Starr}, {Sullivan}, {Velur}, {Walters}, \&
  {Zolkower}}]{ptf2}
{Law}, N.~M., {Kulkarni}, S.~R., {Dekany}, R.~G., {et~al.} 2009, \pasp, 121,
  1395

\bibitem[{{Li} {et~al.}(2000){Li}, {Filippenko}, {Treffers}, {Friedman},
  {Halderson}, {Johnson}, {King}, {Modjaz}, {Papenkova}, {Sato}, \&
  {Shefler}}]{loss}
{Li}, W.~D., {Filippenko}, A.~V., {Treffers}, R.~R., {et~al.} 2000, in American
  Institute of Physics Conference Series, Vol. 522, American Institute of
  Physics Conference Series, ed. S.~S. {Holt} \& W.~W. {Zhang}, 103--106

\bibitem[{{Milne} {et~al.}(2015){Milne}, {Foley}, {Brown}, \&
  {Narayan}}]{milne15}
{Milne}, P.~A., {Foley}, R.~J., {Brown}, P.~J., \& {Narayan}, G. 2015, \apj,
  803, 20

\bibitem[{{Neill} {et~al.}(2011){Neill}, {Sullivan}, {Gal-Yam}, {Quimby},
  {Ofek}, {Wyder}, {Howell}, {Nugent}, {Seibert}, {Martin}, {Overzier},
  {Barlow}, {Foster}, {Friedman}, {Morrissey}, {Neff}, {Schiminovich},
  {Bianchi}, {Donas}, {Heckman}, {Lee}, {Madore}, {Milliard}, {Rich}, \&
  {Szalay}}]{neill11}
{Neill}, J.~D., {Sullivan}, M., {Gal-Yam}, A., {et~al.} 2011, \apj, 727, 15

\bibitem[{{Nicholl} {et~al.}(2014){Nicholl}, {Smartt}, {Jerkstrand}, {Inserra},
  {Anderson}, {Baltay}, {Benetti}, {Chen}, {Elias-Rosa}, {Feindt}, {Fraser},
  {Gal-Yam}, {Hadjiyska}, {Howell}, {Kotak}, {Lawrence}, {Leloudas},
  {Margheim}, {Mattila}, {McCrum}, {McKinnon}, {Mead}, {Nugent}, {Rabinowitz},
  {Rest}, {Smith}, {Sollerman}, {Sullivan}, {Taddia}, {Valenti}, {Walker}, \&
  {Young}}]{nicholl14}
{Nicholl}, M., {Smartt}, S.~J., {Jerkstrand}, A., {et~al.} 2014, \mnras, 444,
  2096

\bibitem[{{Nicholl} {et~al.}(2015){Nicholl}, {Smartt}, {Jerkstrand}, {Inserra},
  {Sim}, {Chen}, {Benetti}, {Fraser}, {Gal-Yam}, {Kankare}, {Maguire}, {Smith},
  {Sullivan}, {Valenti}, {Young}, {Baltay}, {Bauer}, {Baumont}, {Bersier},
  {Botticella}, {Childress}, {Dennefeld}, {Della Valle}, {Elias-Rosa},
  {Feindt}, {Galbany}, {Hadjiyska}, {Le Guillou}, {Leloudas}, {Mazzali},
  {McKinnon}, {Polshaw}, {Rabinowitz}, {Rostami}, {Scalzo}, {Schmidt},
  {Schulze}, {Sollerman}, {Taddia}, \& {Yuan}}]{nicholl15}
---. 2015, ArXiv e-prints, arXiv:1503.03310

\bibitem[{{Pastorello} {et~al.}(2010){Pastorello}, {Smartt}, {Botticella},
  {Maguire}, {Fraser}, {Smith}, {Kotak}, {Magill}, {Valenti}, {Young},
  {Gezari}, {Bresolin}, {Kudritzki}, {Howell}, {Rest}, {Metcalfe}, {Mattila},
  {Kankare}, {Huang}, {Urata}, {Burgett}, {Chambers}, {Dombeck}, {Flewelling},
  {Grav}, {Heasley}, {Hodapp}, {Kaiser}, {Luppino}, {Lupton}, {Magnier},
  {Monet}, {Morgan}, {Onaka}, {Price}, {Rhoads}, {Siegmund}, {Stubbs},
  {Sweeney}, {Tonry}, {Wainscoat}, {Waterson}, {Waters}, \&
  {Wynn-Williams}}]{pastorello10}
{Pastorello}, A., {Smartt}, S.~J., {Botticella}, M.~T., {et~al.} 2010, \apjl,
  724, L16

\bibitem[{Pedregosa {et~al.}(2011)Pedregosa, Varoquaux, Gramfort, Michel,
  Thirion, Grisel, Blondel, Prettenhofer, Weiss, Dubourg, Vanderplas, Passos,
  Cournapeau, Brucher, Perrot, \& Duchesnay}]{sklearn}
Pedregosa, F., Varoquaux, G., Gramfort, A., {et~al.} 2011, Journal of Machine
  Learning Research, 12, 2825

\bibitem[{{Pel} \& {Lub}(2007)}]{walraven}
{Pel}, J.~W., \& {Lub}, J. 2007, in Astronomical Society of the Pacific
  Conference Series, Vol. 364, The Future of Photometric, Spectrophotometric
  and Polarimetric Standardization, ed. C.~{Sterken}, 63

\bibitem[{{Perlmutter} {et~al.}(1999){Perlmutter}, {Aldering}, {Goldhaber},
  {Knop}, {Nugent}, {Castro}, {Deustua}, {Fabbro}, {Goobar}, {Groom}, {Hook},
  {Kim}, {Kim}, {Lee}, {Nunes}, {Pain}, {Pennypacker}, {Quimby}, {Lidman},
  {Ellis}, {Irwin}, {McMahon}, {Ruiz-Lapuente}, {Walton}, {Schaefer}, {Boyle},
  {Filippenko}, {Matheson}, {Fruchter}, {Panagia}, {Newberg}, {Couch}, \&
  {Project}}]{scp99}
{Perlmutter}, S., {Aldering}, G., {Goldhaber}, G., {et~al.} 1999, \apj, 517,
  565

\bibitem[{{Phillips} {et~al.}(2013){Phillips}, {Simon}, {Morrell}, {Burns},
  {Cox}, {Foley}, {Karakas}, {Patat}, {Sternberg}, {Williams}, {Gal-Yam},
  {Hsiao}, {Leonard}, {Persson}, {Stritzinger}, {Thompson}, {Campillay},
  {Contreras}, {Folatelli}, {Freedman}, {Hamuy}, {Roth}, {Shields}, {Suntzeff},
  {Chomiuk}, {Ivans}, {Madore}, {Penprase}, {Perley}, {Pignata}, {Preston}, \&
  {Soderberg}}]{phillips13}
{Phillips}, M.~M., {Simon}, J.~D., {Morrell}, N., {et~al.} 2013, \apj, 779, 38

\bibitem[{{Quimby} {et~al.}(2007){Quimby}, {H{\"o}flich}, \& {Wheeler}}]{qhw07}
{Quimby}, R., {H{\"o}flich}, P., \& {Wheeler}, J.~C. 2007, \apj, 666, 1083

\bibitem[{{Quimby} {et~al.}(2011){Quimby}, {Kulkarni}, {Kasliwal}, {Gal-Yam},
  {Arcavi}, {Sullivan}, {Nugent}, {Thomas}, {Howell}, {Nakar}, {Bildsten},
  {Theissen}, {Law}, {Dekany}, {Rahmer}, {Hale}, {Smith}, {Ofek}, {Zolkower},
  {Velur}, {Walters}, {Henning}, {Bui}, {McKenna}, {Poznanski}, {Cenko}, \&
  {Levitan}}]{quimby11}
{Quimby}, R.~M., {Kulkarni}, S.~R., {Kasliwal}, M.~M., {et~al.} 2011, \nat,
  474, 487

\bibitem[{{Rau} {et~al.}(2009){Rau}, {Kulkarni}, {Law}, {Bloom}, {Ciardi},
  {Djorgovski}, {Fox}, {Gal-Yam}, {Grillmair}, {Kasliwal}, {Nugent}, {Ofek},
  {Quimby}, {Reach}, {Shara}, {Bildsten}, {Cenko}, {Drake}, {Filippenko},
  {Helfand}, {Helou}, {Howell}, {Poznanski}, \& {Sullivan}}]{ptf}
{Rau}, A., {Kulkarni}, S.~R., {Law}, N.~M., {et~al.} 2009, \pasp, 121, 1334

\bibitem[{{Regnault} {et~al.}(2012){Regnault}, {Barrelet}, {Guyonnet},
  {Juramy}, {Rocci}, {Le Guillou}, {Schahman{\`e}che}, \& {Villa}}]{sndice}
{Regnault}, N., {Barrelet}, E., {Guyonnet}, A., {et~al.} 2012, ArXiv e-prints,
  arXiv:1208.6301

\bibitem[{{Riess} {et~al.}(1998){Riess}, {Filippenko}, {Challis},
  {Clocchiatti}, {Diercks}, {Garnavich}, {Gilliland}, {Hogan}, {Jha},
  {Kirshner}, {Leibundgut}, {Phillips}, {Reiss}, {Schmidt}, {Schommer},
  {Smith}, {Spyromilio}, {Stubbs}, {Suntzeff}, \& {Tonry}}]{riess98}
{Riess}, A.~G., {Filippenko}, A.~V., {Challis}, P., {et~al.} 1998, \aj, 116,
  1009

\bibitem[{{Riess} {et~al.}(1999){Riess}, {Kirshner}, {Schmidt}, {Jha},
  {Challis}, {Garnavich}, {Esin}, {Carpenter}, {Grashius}, {Schild}, {Berlind},
  {Huchra}, {Prosser}, {Falco}, {Benson}, {Brice{\~n}o}, {Brown}, {Caldwell},
  {dell'Antonio}, {Filippenko}, {Goodman}, {Grogin}, {Groner}, {Hughes},
  {Green}, {Jansen}, {Kleyna}, {Luu}, {Macri}, {McLeod}, {McLeod}, {McNamara},
  {McLean}, {Milone}, {Mohr}, {Moraru}, {Peng}, {Peters}, {Prestwich},
  {Stanek}, {Szentgyorgyi}, \& {Zhao}}]{cfa}
{Riess}, A.~G., {Kirshner}, R.~P., {Schmidt}, B.~P., {et~al.} 1999, \aj, 117,
  707

\bibitem[{{Schmidt} {et~al.}(1998){Schmidt}, {Suntzeff}, {Phillips},
  {Schommer}, {Clocchiatti}, {Kirshner}, {Garnavich}, {Challis}, {Leibundgut},
  {Spyromilio}, {Riess}, {Filippenko}, {Hamuy}, {Smith}, {Hogan}, {Stubbs},
  {Diercks}, {Reiss}, {Gilliland}, {Tonry}, {Maza}, {Dressler}, {Walsh}, \&
  {Ciardullo}}]{schmidt98}
{Schmidt}, B.~P., {Suntzeff}, N.~B., {Phillips}, M.~M., {et~al.} 1998, \apj,
  507, 46

\bibitem[{{Scolnic} {et~al.}(2014){Scolnic}, {Riess}, {Foley}, {Rest},
  {Rodney}, {Brout}, \& {Jones}}]{scolnic14}
{Scolnic}, D.~M., {Riess}, A.~G., {Foley}, R.~J., {et~al.} 2014, \apj, 780, 37

\bibitem[{{Skrutskie} {et~al.}(2006){Skrutskie}, {Cutri}, {Stiening},
  {Weinberg}, {Schneider}, {Carpenter}, {Beichman}, {Capps}, {Chester},
  {Elias}, {Huchra}, {Liebert}, {Lonsdale}, {Monet}, {Price}, {Seitzer},
  {Jarrett}, {Kirkpatrick}, {Gizis}, {Howard}, {Evans}, {Fowler}, {Fullmer},
  {Hurt}, {Light}, {Kopan}, {Marsh}, {McCallon}, {Tam}, {Van Dyk}, \&
  {Wheelock}}]{2mass}
{Skrutskie}, M.~F., {Cutri}, R.~M., {Stiening}, R., {et~al.} 2006, \aj, 131,
  1163

\bibitem[{{Smartt} {et~al.}(2014){Smartt}, {Valenti}, {Fraser}, {Inserra},
  {Young}, {Sullivan}, {Pastorello}, {Benetti}, {Gal-Yam}, {Knapic},
  {Molinaro}, {Smareglia}, {Smith}, {Taubenberger}, {Yaron}, {Anderson},
  {Ashall}, {Balland}, {Baltay}, {Barbarino}, {Bauer}, {Baumont}, {Bersier},
  {Blagorodnova}, {Bongard}, {Botticella}, {Bufano}, {Bulla}, {Cappellaro},
  {Campbell}, {Cellier-Holzem}, {Chen}, {Childress}, {Clocchiatti},
  {Contreras}, {Dall Ora}, {Danziger}, {de Jaeger}, {De Cia}, {Della Valle},
  {Dennefeld}, {Elias-Rosa}, {Ellman}, {Feindt}, {Fleury}, {Gall},
  {Gonzalez-Gaitan}, {Galbany}, {Morales Garoffolo}, {Greggio}, {Guillou},
  {Hachinger}, {Hadjiyska}, {Hage}, {Hillebrandt}, {Hodgkin}, {Hsiao}, {James},
  {Jerkstrand}, {Kangas}, {Kankare}, {Kotak}, {Kromer}, {Kuncarayakti},
  {Leloudas}, {Lundqvist}, {Lyman}, {Hook}, {Maguire}, {Manulis}, {Margheim},
  {Mattila}, {Maund}, {Mazzali}, {McCrum}, {McKinnon}, {Moreno-Raya},
  {Nicholl}, {Nugent}, {Pain}, {Phillips}, {Pignata}, {Polshaw}, {Pumo},
  {Rabinowitz}, {Reilly}, {Romero-Canizales}, {Scalzo}, {Schmidt}, {Schulze},
  {Sim}, {Sollerman}, {Taddia}, {Tartaglia}, {Terreran}, {Tomasella},
  {Turatto}, {Walker}, {Walton}, {Wyrzykowski}, {Yuan}, \& {Zampieri}}]{pessto}
{Smartt}, S.~J., {Valenti}, S., {Fraser}, M., {et~al.} 2014, ArXiv e-prints,
  arXiv:1411.0299

\bibitem[{{Smith} {et~al.}(2011){Smith}, {Lynn}, {Sullivan}, {Lintott},
  {Nugent}, {Botyanszki}, {Kasliwal}, {Quimby}, {Bamford}, {Fortson},
  {Schawinski}, {Hook}, {Blake}, {Podsiadlowski}, {J{\"o}nsson}, {Gal-Yam},
  {Arcavi}, {Howell}, {Bloom}, {Jacobsen}, {Kulkarni}, {Law}, {Ofek}, \&
  {Walters}}]{snzoo}
{Smith}, A.~M., {Lynn}, S., {Sullivan}, M., {et~al.} 2011, \mnras, 412, 1309

\bibitem[{{Stubbs} {et~al.}(2007){Stubbs}, {Slater}, {Brown}, {Sherman},
  {Smith}, {Tonry}, {Suntzeff}, {Saha}, {Masiero}, \& {Rodney}}]{stubbs06}
{Stubbs}, C.~W., {Slater}, S.~K., {Brown}, Y.~J., {et~al.} 2007, in
  Astronomical Society of the Pacific Conference Series, Vol. 364, The Future
  of Photometric, Spectrophotometric and Polarimetric Standardization, ed.
  C.~{Sterken}, 373

\bibitem[{{Sullivan} {et~al.}(2009){Sullivan}, {Ellis}, {Howell}, {Riess},
  {Nugent}, \& {Gal-Yam}}]{sullivan09}
{Sullivan}, M., {Ellis}, R.~S., {Howell}, D.~A., {et~al.} 2009, \apjl, 693, L76

\bibitem[{{Sullivan} {et~al.}(2010){Sullivan}, {Conley}, {Howell}, {Neill},
  {Astier}, {Balland}, {Basa}, {Carlberg}, {Fouchez}, {Guy}, {Hardin}, {Hook},
  {Pain}, {Palanque-Delabrouille}, {Perrett}, {Pritchet}, {Regnault}, {Rich},
  {Ruhlmann-Kleider}, {Baumont}, {Hsiao}, {Kronborg}, {Lidman}, {Perlmutter},
  \& {Walker}}]{sullivan10}
{Sullivan}, M., {Conley}, A., {Howell}, D.~A., {et~al.} 2010, \mnras, 406, 782

\bibitem[{{Sullivan} {et~al.}(2011){Sullivan}, {Guy}, {Conley}, {Regnault},
  {Astier}, {Balland}, {Basa}, {Carlberg}, {Fouchez}, {Hardin}, {Hook},
  {Howell}, {Pain}, {Palanque-Delabrouille}, {Perrett}, {Pritchet}, {Rich},
  {Ruhlmann-Kleider}, {Balam}, {Baumont}, {Ellis}, {Fabbro}, {Fakhouri},
  {Fourmanoit}, {Gonz{\'a}lez-Gait{\'a}n}, {Graham}, {Hudson}, {Hsiao},
  {Kronborg}, {Lidman}, {Mourao}, {Neill}, {Perlmutter}, {Ripoche}, {Suzuki},
  \& {Walker}}]{sullivan11}
{Sullivan}, M., {Guy}, J., {Conley}, A., {et~al.} 2011, \apj, 737, 102

\bibitem[{{Wang} {et~al.}(2009){Wang}, {Filippenko}, {Ganeshalingam}, {Li},
  {Silverman}, {Wang}, {Chornock}, {Foley}, {Gates}, {Macomber}, {Serduke},
  {Steele}, \& {Wong}}]{wang09}
{Wang}, X., {Filippenko}, A.~V., {Ganeshalingam}, M., {et~al.} 2009, \apjl,
  699, L139

\bibitem[{{York} {et~al.}(2000){York}, {Adelman}, {Anderson}, {Anderson},
  {Annis}, {Bahcall}, {Bakken}, {Barkhouser}, {Bastian}, {Berman}, {Boroski},
  {Bracker}, {Briegel}, {Briggs}, {Brinkmann}, {Brunner}, {Burles}, {Carey},
  {Carr}, {Castander}, {Chen}, {Colestock}, {Connolly}, {Crocker}, {Csabai},
  {Czarapata}, {Davis}, {Doi}, {Dombeck}, {Eisenstein}, {Ellman}, {Elms},
  {Evans}, {Fan}, {Federwitz}, {Fiscelli}, {Friedman}, {Frieman}, {Fukugita},
  {Gillespie}, {Gunn}, {Gurbani}, {de Haas}, {Haldeman}, {Harris}, {Hayes},
  {Heckman}, {Hennessy}, {Hindsley}, {Holm}, {Holmgren}, {Huang}, {Hull},
  {Husby}, {Ichikawa}, {Ichikawa}, {Ivezi{\'c}}, {Kent}, {Kim}, {Kinney},
  {Klaene}, {Kleinman}, {Kleinman}, {Knapp}, {Korienek}, {Kron}, {Kunszt},
  {Lamb}, {Lee}, {Leger}, {Limmongkol}, {Lindenmeyer}, {Long}, {Loomis},
  {Loveday}, {Lucinio}, {Lupton}, {MacKinnon}, {Mannery}, {Mantsch}, {Margon},
  {McGehee}, {McKay}, {Meiksin}, {Merelli}, {Monet}, {Munn}, {Narayanan},
  {Nash}, {Neilsen}, {Neswold}, {Newberg}, {Nichol}, {Nicinski}, {Nonino},
  {Okada}, {Okamura}, {Ostriker}, {Owen}, {Pauls}, {Peoples}, {Peterson},
  {Petravick}, {Pier}, {Pope}, {Pordes}, {Prosapio}, {Rechenmacher}, {Quinn},
  {Richards}, {Richmond}, {Rivetta}, {Rockosi}, {Ruthmansdorfer}, {Sandford},
  {Schlegel}, {Schneider}, {Sekiguchi}, {Sergey}, {Shimasaku}, {Siegmund},
  {Smee}, {Smith}, {Snedden}, {Stone}, {Stoughton}, {Strauss}, {Stubbs},
  {SubbaRao}, {Szalay}, {Szapudi}, {Szokoly}, {Thakar}, {Tremonti}, {Tucker},
  {Uomoto}, {Vanden Berk}, {Vogeley}, {Waddell}, {Wang}, {Watanabe},
  {Weinberg}, {Yanny}, {Yasuda}, \& {SDSS Collaboration}}]{sdss}
{York}, D.~G., {Adelman}, J., {Anderson}, Jr., J.~E., {et~al.} 2000, \aj, 120,
  1579

\bibitem[{{Yuan} {et~al.}(2013){Yuan}, {Kobayashi}, {Schmidt}, {Podsiadlowski},
  {Sim}, \& {Scalzo}}]{yuan13}
{Yuan}, F., {Kobayashi}, C., {Schmidt}, B.~P., {et~al.} 2013, \mnras, 432, 1680

\bibitem[{{Zacharias} {et~al.}(2000){Zacharias}, {Urban}, {Zacharias}, {Hall},
  {Wycoff}, {Rafferty}, {Germain}, {Holdenried}, {Pohlman}, {Gauss}, {Monet},
  \& {Winter}}]{ucac2}
{Zacharias}, N., {Urban}, S.~E., {Zacharias}, M.~I., {et~al.} 2000, \aj, 120,
  2131

\end{thebibliography}

\end{document}